\documentclass[a4paper,11pt]{article}
\usepackage{jheppub}
\usepackage[utf8]{inputenc}

\usepackage[vcentermath]{youngtab}
\usepackage{pdflscape}
\usepackage{tikz}
\usetikzlibrary{positioning}
\usetikzlibrary{arrows}
\usetikzlibrary{arrows.meta}

\usetikzlibrary{arrows,fit,decorations.pathreplacing, shapes.misc,decorations.pathmorphing}
\usepackage[capitalise]{cleveref}
\usepackage{xcolor}
\usepackage{soul}
\usepackage{colortbl}
\definecolor{Gray}{gray}{0.9}
\usepackage{multirow}
\usepackage[colorinlistoftodos]{todonotes}
\usepackage{amsthm}
\usepackage{amsmath}
\usepackage{amsthm}
\usepackage{changepage}
\usepackage{tabularx}
\usepackage{tikz}
\usepackage{multirow}
\usepackage{longtable}
\usepackage{multirow}
\usepackage{amsmath}
\usepackage{amssymb}
\usepackage{mathtools}
\usepackage{adjustbox}
\usetikzlibrary{shapes,arrows}
\usepackage[english]{babel}
\newcolumntype{P}[1]{>{\centering\arraybackslash}p{#1}}

\usepackage{color}
\usepackage{subcaption}
\usepackage{gensymb}
\usepackage{pdflscape}
\usepackage{booktabs}

\usepackage{tikz}
\usetikzlibrary{decorations.markings}
\usetikzlibrary{positioning}
\usetikzlibrary{arrows}
\usetikzlibrary{arrows.meta}
\usetikzlibrary{shapes}
\usetikzlibrary{chains}
\usetikzlibrary{arrows,fit,decorations.pathreplacing, shapes.misc,decorations.pathmorphing}
\tikzstyle{every picture}+=[remember picture]
\tikzstyle{na} = [baseline=-.5ex]

\tikzset{flavour/.style={regular polygon,regular polygon sides=4,inner sep=2.5pt, draw}}
\tikzset{gauge/.style={circle, draw,inner sep=2.5pt}}
\tikzset{gaugeb/.style={circle, draw,fill=black,inner sep=2.5pt}}
\tikzset{bd/.style={circle,fill=black, draw,inner sep=0.5pt}}
\tikzset{hasse/.style={circle, fill,inner sep=2pt}}
\tikzset{sev/.style={inner sep=1mm,draw=none,fill=white,minimum size=4mm,circle, draw}}

\usetikzlibrary{shapes.misc}
\tikzset{gauge1/.style={draw=none,minimum size=0.6cm,fill=white,circle, draw}}
\tikzset{gauge3/.style={draw=none,minimum size=0.4cm,fill=white,circle, draw}}
\tikzset{crosses/.style={cross out, draw=black, minimum size=0.3cm, inner sep=0pt, outer sep=0pt},
cross/.default={1pt}}
\tikzset{blank/.style={draw=none,minimum size=0.4cm,fill=none,circle, draw}}
\tikzset{flavour2/.style={draw=none,minimum size=0.4cm,fill=white,regular polygon sides=4,draw}}
\tikzset{flavourBlue/.style={draw=none,minimum size=0.4cm,fill=blue,regular polygon sides=4,draw}}
\tikzset{flavorBlue/.style={draw=none,minimum size=0.4cm,fill=blue,regular polygon sides=4,draw}}
\tikzset{flavourRed/.style={draw=none,minimum size=0.4cm,fill=red,regular polygon sides=4,draw}}
\tikzset{flavorRed/.style={draw=none,minimum size=0.4cm,fill=red,regular polygon sides=4,draw}}
\tikzset{none/.style={draw=none}}
\tikzset{redgauge/.style={draw=none,minimum size=0.4cm,fill=red,circle, draw}}
\tikzset{miniU/.style={draw=none,minimum size=0.1cm,fill=red,circle, draw}}
\tikzset{smallgauge1/.style={draw=none,minimum size=0.1cm,fill=white,circle, draw}}
\tikzset{miniBlue/.style={draw=none,minimum size=0.1cm,fill=blue,circle, draw}}
\tikzset{gauge2/.style={draw=none,minimum size=0.35mm,fill=red,circle, draw}}
\tikzset{bluegauge/.style={draw=none,minimum size=0.4cm,fill=blue,circle, draw}}
\tikzset{flavour1/.style={draw=none,minimum size=0.35mm,fill=blue, regular polygon,regular polygon sides=4,draw}}
\tikzset{flavour0/.style={draw=none,minimum size=0.35mm,fill=white, regular polygon,regular polygon sides=4,draw}}
\tikzset{smalldot/.style={draw=none,minimum size=0.1mm,fill=black, circle,draw}}
\tikzset{dotsize/.style={circle,fill,inner sep=1.5pt,draw}}
\tikzset{doubleguys/.style={double, double distance = 3pt}}
\tikzset{tripleguys/.style={triple}}
\tikzset{new edge style 1/.style={dashed}}
\tikzset{thickline/.style={line width=0.06cm}}
\tikzset{brace/.style={decorate,decoration={brace,amplitude=10pt}}}
\pgfdeclarelayer{edgelayer}
\pgfdeclarelayer{nodelayer}
\usetikzlibrary{decorations.pathreplacing}
\pgfsetlayers{edgelayer,nodelayer,main}

\newcommand{\req}{\rotatebox[origin=c]{90}{=}}
\newcommand{\doubleslash}{\mathbin{/\mkern-6mu/}}
\newcommand{\tripleslash}{\mathbin{/\mkern-6mu/\mkern-6mu/}}

\newcommand{\athree}[3]{\raisebox{-.5\height}{\scalebox{.8}{\begin{tikzpicture}
	\node (g1) [gauge, fill=#1] at (0,0) {};
	\node (g2) [gauge, fill=#2] at (1,0) {};
	\node (g3) [gauge, fill=#3] at (2,0) {};
	\draw (g1)--(g2)--(g3);
	\end{tikzpicture}}}}

\newcommand{\dfourquiver}[4]{\scalebox{.8}{\begin{tikzpicture}
	\node (g1) [gauge, fill=#1] at (0,0) {};
	\node (g2) [gauge, fill=#2] at (1,0) {};
	\node (g3) [gauge, fill=#3] at (1.5,.86) {};
	\node (g4) [gauge, fill=#4] at (1.5,-.86) {};
	\draw (g1)--(g2)--(g3) (g2)--(g4);
	\end{tikzpicture}}}

\preprint{Imperial/TP/21/AH/8}

\title{Partial Implosions and Quivers}

\author[a,b]{Antoine Bourget}
\author[c]{, Andrew Dancer}
\author[d]{, Julius F. Grimminger}
\author[d]{, Amihay Hanany}
\author[d]{and Zhenghao Zhong}
\affiliation[a]{Université Paris-Saclay, CNRS, CEA, Institut de physique théorique, 91191, Gif-sur-Yvette, France}
\affiliation[b]{Laboratoire de Physique de l’Ecole normale supérieure, ENS, Université PSL, CNRS, Sorbonne
Université, Université de Paris, F-75005 Paris, France}
\affiliation[c]{Jesus College, Oxford University, OX1 3DW, UK}
\affiliation[d]{Theoretical Physics Group, The Blackett Laboratory, Imperial College London, Prince Consort Road
London, SW7 2AZ, UK}
\emailAdd{antoine.bourget@polytechnique.org}
\emailAdd{dancer@maths.ox.ac.uk}
\emailAdd{julius.grimminger17@imperial.ac.uk}
\emailAdd{a.hanany@imperial.ac.uk}
\emailAdd{zhenghao.zhong14@imperial.ac.uk}

\abstract{We propose magnetic quivers for partial implosion spaces. Such partial implosions involve a choice of parabolic subgroup, with the Borel subgroup corresponding to the standard implosion. In the subregular case we test the conjecture by verifying that reduction by the Levi group gives the appropriate nilpotent orbit closure. In the case of a parabolic corresponding to a hook diagram we are also able to carry out this verification provided we work at nonzero Fayet-Iliopoulos parameters.}

\begin{document} 

\maketitle

\section{Introduction}

In this paper we provide a physical realization of the geometric concept of \emph{partial implosion} for hyperK\"ahler manifolds, using moduli spaces of supersymmetric quiver gauge theories. Implosion can be seen as an abelianization procedure: given a hyperK\"ahler manifold $M$ with an action of a rank $n$ complex Lie group $G$, its implosion $M_{\textrm{impl}}$ is a lower dimensional hyperK\"ahler manifold with an action of the torus $(\mathbb{C}^{\ast})^n$. \emph{Partial} implosion corresponds to an intermediate case, where the partially imploded manifold $M_{P,\textrm{impl}}$ has an action of a not necessarily Abelian subgroup $L_P$ of $G$. Possible partial implosions are labelled by parabolic subgroups $P$, or equivalently subsets of simple roots of $G$, and $L_P$ is the corresponding Levi subgroup. 

In this paper we focus on the special class of manifolds $M = T^{\ast} G$ for $G = \mathrm{SL}_n := \mathrm{SL}(n,\mathbb{C})$ and $G = \mathrm{SO}_{2n} := \mathrm{SO}(2n,\mathbb{C})$. This is a natural restriction, as the (partial) implosion of any space can be reduced to the (partial) implosion of these spaces -- for that reason, the spaces $(T^{\ast} G)_{P,\textrm{impl}}$ are called \emph{universal partial implosions}. Our main result is a conjecture for magnetic quivers of these universal partial implosions, i.e. we propose quivers $\mathsf{Q}_P$ such that
\begin{equation}
\label{conjecture}
    \mathcal{C} (\mathsf{Q}_P) = (T^{\ast} G)_{P,\textrm{impl}} \, , 
\end{equation}
where $\mathcal{C} (\mathsf{Q}_P)$ is the Coulomb branch of the 3d $\mathcal{N}=4$ theory defined by $\mathsf{Q}_P$. The quivers $\mathsf{Q}_P$ are presented in equations (\ref{implodedQuiver}) and (\ref{partialImplosionSO}). This generalizes \cite{DHK,Bourget:2021zyc} where magnetic quivers for \emph{full} implosion spaces were considered. The conjecture (\ref{conjecture}) is backed by dimension considerations as well as explicit computations for families of parabolics $P$, corresponding to hook partitions for $\mathrm{SL}_n$ and the so-called $E_6$ family (defined in Section \ref{sec:E6family}) for $\mathrm{SO}_{2n}$. 

The paper is organised as follows. In Sections \ref{sec:phys} and \ref{sec:math} we review some physical and mathematical background. In Section \ref{sec:unitary} we study universal partial implosions for $G = \mathrm{SL}_n$, and in Section \ref{sec:orthosymplectic} for $G = \mathrm{SO}_{2n}$.  

\subsection{Physical background}
\label{sec:phys}

To a 3d $\mathcal{N}=4$ quiver gauge theory\footnote{In a slight abuse of notation we refer to a quiver, and the 3d $\mathcal{N}=4$ theory it defines, with the same symbol. In fact, not every 3d $\mathcal{N}=4$ theory has a quiver description, but in this paper we only consider those which do.} $\mathsf{Q}$ we can associate two hyperK\"ahler varieties, the Coulomb branch $\mathcal{C}(\mathsf{Q})$ and the Higgs branch $\mathcal{H}(\mathsf{Q})$. While the Higgs branch is simply a hyperK\"ahler quotient \cite{Hitchin:1986ea}, the Coulomb branch is a moduli space of dressed monopole operators \cite{Cremonesi:2013lqa,Bullimore:2015lsa}, with a mathematical definition in its own right \cite{Nakajima:2015txa,Braverman:2016wma}. Its Hilbert series is computed using the so-called monopole formula \cite{Cremonesi:2013lqa}.

\paragraph{Electric and magnetic quiver.} Given a hyperK\"ahler variety $\mathcal{V}$ we call $\mathsf{Q}_e$ an \emph{electric quiver} for $\mathcal{V}$, if $\mathcal{V}=\mathcal{H}(\mathsf{Q}_e)$, and $\mathsf{Q}_m$ a \emph{magnetic quiver} for $\mathcal{V}$, if $\mathcal{V}=\mathcal{C}(\mathsf{Q}_m)$. A variety $\mathcal{V}$ may have many electric and magnetic quivers.

Throughout the paper all our quivers contain only unitary, special unitary, (special) orthogonal and symplectic nodes, as well as hypermultiplets in (bi-)fundamental representations.

\paragraph{Balance.} Given a quiver with gauge nodes\footnote{In another abuse of notation, we identify a gauge node with its associated gauge group.} $G_i$, for each $G_i$ call $N_f$ the number of hypermultiplets in the fundamental representation of $G_i$. E.g.\ for SQCD we have
\begin{equation}
\label{eq:SQCDquivers}
	\raisebox{-.5\height}{\begin{tikzpicture}
		\node[gauge3,label=below:{U$(N_c)$}] (1) at (0,0) {};
		\node[flavour2,label=above:{SU$(N_f)$}] (2) at (0,1) {};
		\draw (1)--(2);
	\end{tikzpicture}
	\qquad
	\begin{tikzpicture}
		\node[bluegauge,label=below:{USp$(N_c)$}] (1) at (0,0) {};
		\node[flavourRed,label=above:{O$(N_f)$}] (2) at (0,1) {};
		\draw (1)--(2);
	\end{tikzpicture}
	\qquad
	\begin{tikzpicture}
		\node[redgauge,label=below:{(S)O$(N_c)$}] (1) at (0,0) {};
		\node[flavourBlue,label=above:{USp$(N_f)$}] (2) at (0,1) {};
		\draw (1)--(2);
	\end{tikzpicture}}\;.
\end{equation}

A useful notion is the so called \emph{balance} $b_{G_i}$ of the gauge node $G_i$.
We define
\begin{equation}
	b_{G_i}=\begin{cases}
		N_f-2N_c & \textnormal{if }G_i=\mathrm{U}(N_c)\\
		N_f-2N_c-2 & \textnormal{if }G_i=\mathrm{USp}(N_c)\\
		N_f-2N_c+2 & \textnormal{if }G_i=\mathrm{(S)O}(N_c)\\
	\end{cases}\;.
\end{equation}
We call a node \emph{underbalanced} if $b<0$, \emph{minimally unbalanced} if $b=-1$, \emph{balanced} if $b=0$ and \emph{overbalanced} if $b>0$. When nodes in the quiver have $b<-1$ then the monopole formula fails to compute the Coulomb branch Hilbert series, and the Coulomb branch is generally not a cone. When $b\geq-1$ for all nodes, then the monopole formula works for all quivers in this paper.

\paragraph{Coulomb branch global symmetry.} For every U$(N_c)$ node there is a U$(1)$ factor in the global symmetry of the Coulomb branch. For orthosymplectic nodes there is only a discrete factor. If there are balanced gauge nodes, then the global symmetry of the Coulomb branch enhances. One can perform a hyperK\"ahler quotient of the Coulomb branch by a subgroup of this global symmetry. Physically this corresponds to a gauging of the global symmetry. However it is not always easy to implement this hyperK\"ahler quotient, as there are issues of incomplete Higgsing.

\paragraph{(In)complete Higgsing.} For a given quiver, one can ask whether the gauge group can be completely broken. This cannot be achieved on the Coulomb branch as the gauge group here is at most broken to its maximal torus. On a generic point on the Higgs branch the gauge group may be: 1) completely broken -- this is called \emph{complete} Higgsing, or 2) broken to a subgroup -- this is called \emph{incomplete} Higgsing. Mathematically incomplete Higgsing means, that the group by which one performs a hyperK\"ahler quotient acts non-freely.
For the SQCD theories in \eqref{eq:SQCDquivers} we have incomplete Higgsing, if $N_f<2N_c$.

When there is incomplete Higgsing several complications can arise, and in particular the Higgs branch Hilbert series is difficult to compute. Furthermore Fayet-Iliopoulos parameters act in an intricate way.

\paragraph{Fayet-Iliopoulos parameters.} Given a quiver as described above there are several deformation parameters one can turn on. For a U($N_c$) node there is a Fayet-Iliopoulos (FI) parameter which can be turned on in the Lagrangian. For a USp($N_c)$ or a (S)O($N_c$) node there is no such FI term in the Lagrangian. However there is a conjectured deformation at the fixed point in the IR, called \emph{hidden} FI parameter.

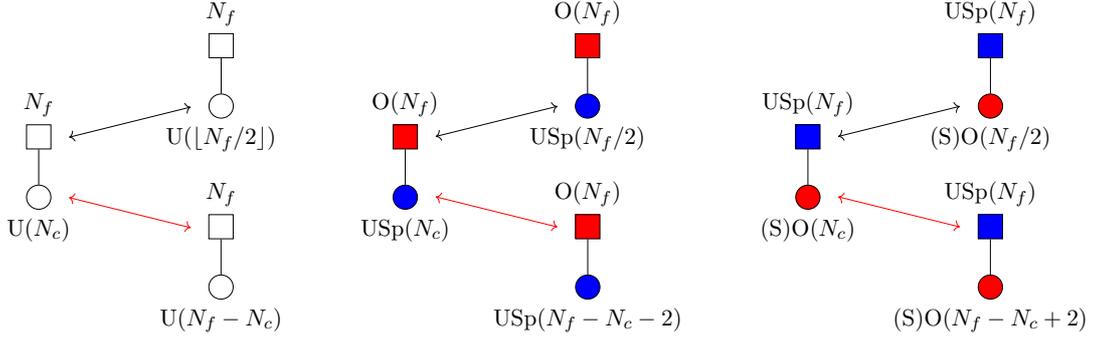
\begin{figure}
    \centering
       \scalebox{.8}{
    \begin{tikzpicture}
        \node (a) at (0,0) {$\begin{tikzpicture}
            \node[gauge3,label=below:{U$(N_c)$}] (a1) at (0,0) {};
	        \node[flavour2,label=above:{$N_f$}] (a2) at (0,1) {};
		    \draw (a1)--(a2);
        \end{tikzpicture}$};
        \node (b) at (3,1.5) {$\begin{tikzpicture}
            \node[gauge3,label=below:{U$(\left \lfloor{N_f/2}\right \rfloor)$}] (b1) at (0,0) {};
	        \node[flavour2,label=above:{$N_f$}] (b2) at (0,1) {};
		    \draw (b1)--(b2);
        \end{tikzpicture}$};
        \node (c) at (3,-1.5) {$\begin{tikzpicture}
            \node[gauge3,label=below:{U$(N_f-N_c)$}] (c1) at (0,0) {};
	        \node[flavour2,label=above:{$N_f$}] (c2) at (0,1) {};
		    \draw (c1)--(c2);
        \end{tikzpicture}$};
        \draw[<->] (0.5,0.5)--(2.5,1);
        \draw[red,<->] (0.5,-0.5)--(2.5,-1);
	\end{tikzpicture}\qquad
    \begin{tikzpicture}
        \node (a) at (0,0) {$\begin{tikzpicture}
            \node[bluegauge,label=below:{USp$(N_c)$}] (a1) at (0,0) {};
	        \node[flavourRed,label=above:{O($N_f$)}] (a2) at (0,1) {};
		    \draw (a1)--(a2);
        \end{tikzpicture}$};
        \node (b) at (3,1.5) {$\begin{tikzpicture}
            \node[bluegauge,label=below:{USp$(N_f/2)$}] (b1) at (0,0) {};
	        \node[flavourRed,label=above:{O($N_f$)}] (b2) at (0,1) {};
		    \draw (b1)--(b2);
        \end{tikzpicture}$};
        \node (c) at (3,-1.5) {$\begin{tikzpicture}
            \node[bluegauge,label=below:{USp$(N_f-N_c-2)$}] (c1) at (0,0) {};
	        \node[flavourRed,label=above:{O($N_f$)}] (c2) at (0,1) {};
		    \draw (c1)--(c2);
        \end{tikzpicture}$};
        \draw[<->] (0.5,0.5)--(2.5,1);
        \draw[red,<->] (0.5,-0.5)--(2.5,-1);
	\end{tikzpicture}\qquad
    \begin{tikzpicture}
        \node (a) at (0,0) {$\begin{tikzpicture}
            \node[redgauge,label=below:{(S)O$(N_c)$}] (a1) at (0,0) {};
	        \node[flavourBlue,label=above:{USp($N_f$)}] (a2) at (0,1) {};
		    \draw (a1)--(a2);
        \end{tikzpicture}$};
        \node (b) at (3,1.5) {$\begin{tikzpicture}
            \node[redgauge,label=below:{(S)O$(N_f/2)$}] (b1) at (0,0) {};
	        \node[flavourBlue,label=above:{USp($N_f$)}] (b2) at (0,1) {};
		    \draw (b1)--(b2);
        \end{tikzpicture}$};
        \node (c) at (3,-1.5) {$\begin{tikzpicture}
            \node[redgauge,label=below:{(S)O$(N_f-N_c+2)$}] (c1) at (0,0) {};
	        \node[flavourBlue,label=above:{USp($N_f$)}] (c2) at (0,1) {};
		    \draw (c1)--(c2);
        \end{tikzpicture}$};
        \draw[<->] (0.5,0.5)--(2.5,1);
        \draw[red,<->] (0.5,-0.5)--(2.5,-1);
	\end{tikzpicture}}
    \caption{Effect of FI parameters on the Higgs branch of SQCD theories with incomplete Higgsing ($N_f<2N_c$). Black arrows (respectively red arrows) indicate equality of Higgs branches with zero (resp. non-zero) FI. If $N_f-N_c+\epsilon<0$, where $\epsilon=0$ for U$(N_c)$, $\epsilon=-2$ for USp$(N_c)$, and $\epsilon=2$ for (S)O$(N_c)$, there is no vacuum preserving 8 supercharges with non-zero FI.}
    \label{fig:(Seiberg)Duals}
\end{figure}

Let us consider SQCD theories. When there is complete Higgsing, the FI parameter lifts the Coulomb branch of the theory and resolves / deforms the Higgs branch. When there is incomplete Higgsing the FI parameter has a more violent effect, described in Figure \ref{fig:(Seiberg)Duals}. If the FI parameter is zero, then the Higgs branch of the theory is equal to the Higgs branch of a different SQCD theory denoted by a black arrow in Figure \ref{fig:(Seiberg)Duals}. If the FI parameter is non-zero then the Higgs branch is not the resolution / deformation of the the Higgs branch for $\textrm{FI}=0$, it is the resolution / deformation of the Higgs branch of a different SQCD theory with smaller dimension, denoted by a red arrow in Figure \ref{fig:(Seiberg)Duals}. For the unitary case this is discussed in detail in \cite[App. B]{Bourget:2021jwo} based on \cite{Yaakov:2013fza,Assel:2017jgo}. For the orthosymplectic case the $\textrm{FI}=0$ case can be obtained by looking at the partial Higgs mechanism, see e.g.\ \cite{Bourget:2019aer}, the $\textrm{FI}\neq0$ case is given in \cite{Seiberg:1994pq,Aharony:1997gp,Kapustin:2011gh} but in a slightly different context of Seiberg duality of orthosymplectic SQCD theories in 3d $\mathcal{N}=2$.

\subsection{Mathematical background}
\label{sec:math}

\begin{figure}
\makebox[\textwidth][c]{\begin{tikzpicture}
\node (aa) at (0,10) {$\begin{array}{c}
 T^\ast K  \\ \req \\ \color{red}{K_\mathbb{C}}
\end{array}  $};
\node (bb) at (0,7) {$\begin{array}{c}
(T^\ast K)_{\mathrm{impl}} \\ \req \\ \color{red}{K_\mathbb{C} \doubleslash_{\mathrm{GIT}} N}
\end{array}  $};
\node at (0,6) {$\dim_{\mathbb{R}} = 2(r + \delta)$};
\node at (-2,10) {$\curvearrowright$};
\node at (2,10) {$\curvearrowleft$};
\node at (-2,7) {$\curvearrowright$};
\node at (2,7) {$\curvearrowleft$};
\node at (-3,10) {$\color{olive}{K}$};
\node at (3,10) {$\color{purple}{K}$};
\node at (6,10) {$\color{olive}{K}$};
\node at (7,10) {$\curvearrowright$};
\node at (9,10) {$T^\ast K\doubleslash^{s}_{\lambda} \color{purple}{K}$};
\node at (-3,7) {$\color{olive}{K}$};
\node at (3,7) {$\color{purple}{T}$};
\node at (6,7) {$\color{olive}{K}$};
\node at (7,7) {$\curvearrowright$};
\node at (9,7) {$(T^\ast K)_{\mathrm{impl}}\doubleslash^{s}_{\lambda} {\color{purple}{T}} = \mathcal{O}_{\lambda}$};
\draw (8.9,8)--(8.9,9);
\draw (9,8)--(9,9);
\draw (4.5,12)--(4.5,6);
\draw[->,thick] (0,9)--(0,8);
\node at (1,8.5) {Implosion};
\node at (0,11.5) {\textsc{Symplectic Implosion}};
\node at (8,11.5) {\textsc{Reduced Spaces}};
\end{tikzpicture}}
\caption{Universal symplectic implosion. Here $K$ is a rank $r$ compact simple Lie group with maximal torus $T$, and $N$ is a maximal unipotent subgroup of the complex group $K_{\mathbb{C}}$. $\delta$ is the number of positive roots. $\doubleslash^{s}_{\lambda}$ denotes the symplectic reduction at level $\lambda \in \mathfrak{t}^{\ast}$ and $\mathcal{O}_{\lambda}$ is the orbit of $K$ through $\lambda$. }
\label{fig:implosionSummary1}
\end{figure}
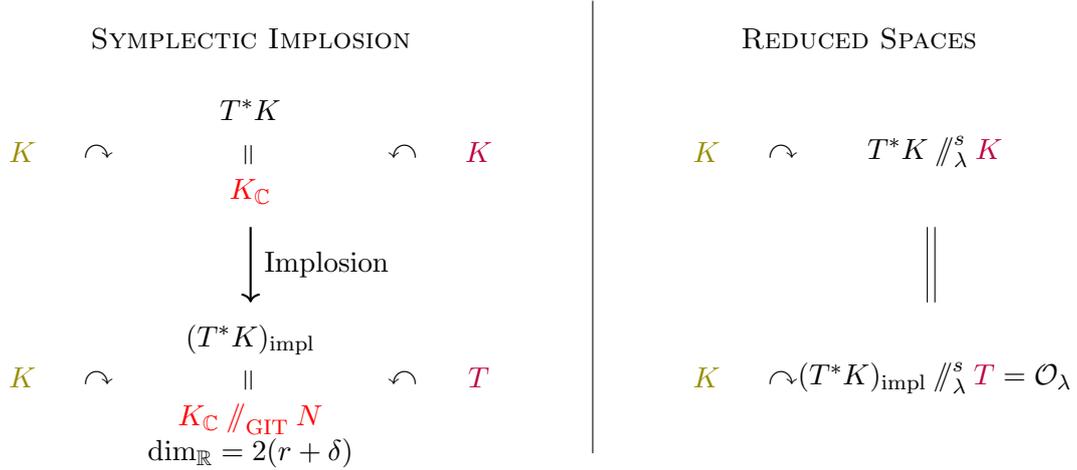

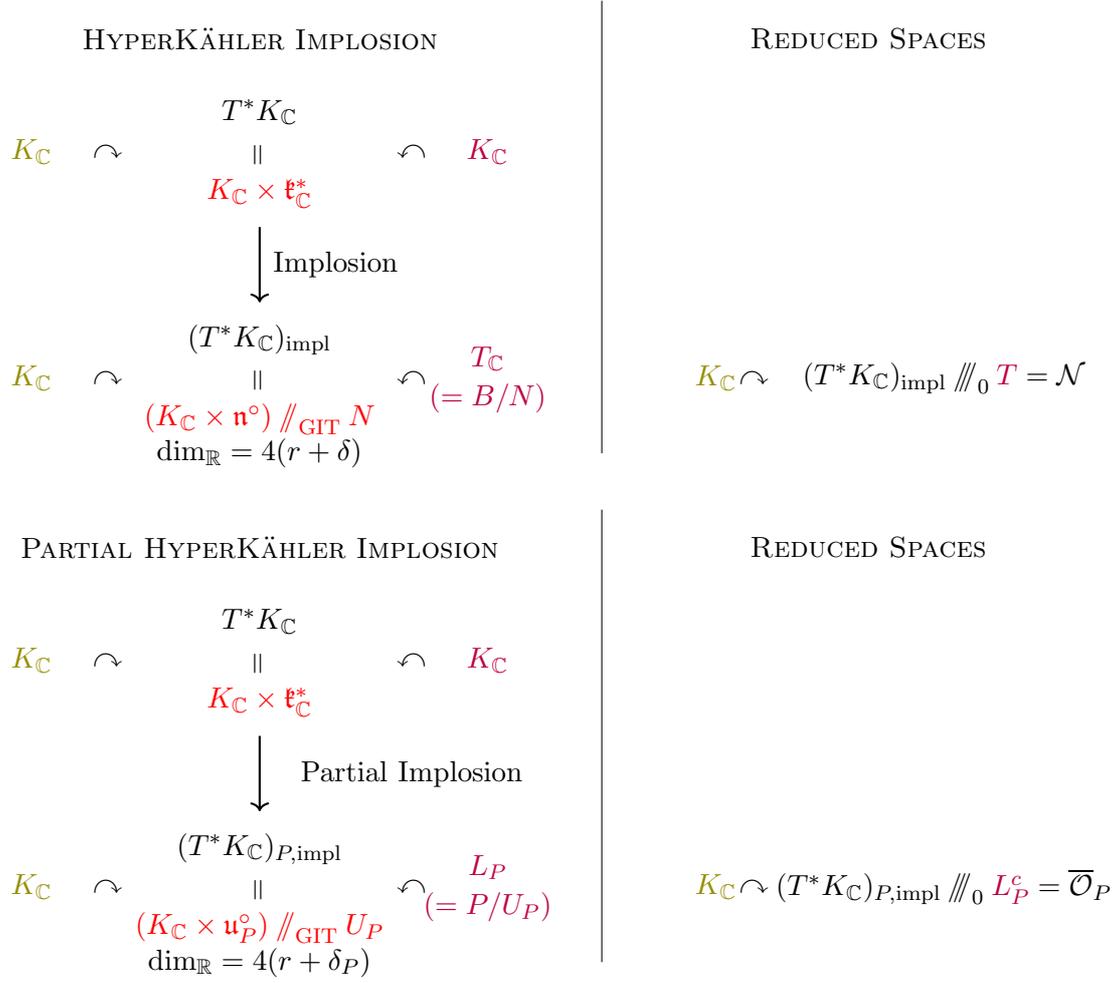
\begin{figure}
\makebox[\textwidth][c]{\begin{tikzpicture}
\node at (0,10) {$\begin{array}{c}
 T^\ast K_{\mathbb{C}}  \\ \req \\ \color{red}{K_\mathbb{C} \times \mathfrak{k}_{\mathbb{C}}^\ast}
\end{array}  $};
\node at (0,7) {$\begin{array}{c}
(T^\ast K_{\mathbb{C}})_{\mathrm{impl}} \\ \req \\ {\color{red}{(K_\mathbb{C} \times \mathfrak{n}^{\circ}}) \doubleslash_{\mathrm{GIT}} N}
\end{array}  $};
\node at (0,6) {$\dim_{\mathbb{R}} = 4(r + \delta)$};
\node at (-2,10) {$\curvearrowright$};
\node at (2,10) {$\curvearrowleft$};
\node at (-2,7) {$\curvearrowright$};
\node at (2,7) {$\curvearrowleft$};
\node at (-3,10) {$\color{olive}{K_{\mathbb{C}}}$};
\node at (3,10) {$\color{purple}{K_{\mathbb{C}}}$};
\node at (-3,7) {$\color{olive}{K_{\mathbb{C}}}$};
\node at (3,7) {$\color{purple}{\begin{array}{c}
T_{\mathbb{C}} \\ (=B/N)
\end{array}  }$};
\node at (6,7) {$\color{olive}{K_{\mathbb{C}}}$};
\node at (6.5,7) {$\curvearrowright$};
\node at (9,7) {$(T^\ast K_{\mathbb{C}})_{\mathrm{impl}} \tripleslash_{0} {\color{purple}{T}} = \mathcal{N}$};
\draw (4.5,12)--(4.5,6);
\draw[->,thick] (0,9)--(0,8);
\node at (1,8.5) {Implosion};
\node at (0,11.5) {\textsc{HyperK\"ahler Implosion}};
\node at (8,11.5) {\textsc{Reduced Spaces}};
\node at (9,9) {\textcolor{white}{$(T^\ast K_{\mathbb{C}})_{P,\mathrm{impl}} \tripleslash_{0} {L_P} = \overline{\mathcal{O}}_P$}};
\end{tikzpicture}}

\vspace{1em}

\makebox[\textwidth][c]{\begin{tikzpicture}
\node at (0,10) {$\begin{array}{c}
 T^\ast K_{\mathbb{C}}  \\ \req \\ \color{red}{K_\mathbb{C} \times \mathfrak{k}_{\mathbb{C}}^\ast}
\end{array}  $};
\node at (0,7) {$\begin{array}{c}
(T^\ast K_{\mathbb{C}})_{P,\mathrm{impl}} \\ \req \\ {\color{red}{(K_\mathbb{C} \times \mathfrak{u}_P^{\circ}}) \doubleslash_{\mathrm{GIT}} U_P}
\end{array}  $};
\node at (0,6) {$\dim_{\mathbb{R}} = 4(r + \delta_P)$};
\node at (-2,10) {$\curvearrowright$};
\node at (2,10) {$\curvearrowleft$};
\node at (-2,7) {$\curvearrowright$};
\node at (2,7) {$\curvearrowleft$};
\node at (-3,10) {$\color{olive}{K_{\mathbb{C}}}$};
\node at (3,10) {$\color{purple}{K_{\mathbb{C}}}$};
\node at (-3,7) {$\color{olive}{K_{\mathbb{C}}}$};
\node at (3,7) {$\color{purple}{\begin{array}{c}
L_P \\ (=P/U_P)
\end{array}  }$};
\node at (6,7) {$\color{olive}{K_{\mathbb{C}}}$};
\node at (6.5,7) {$\curvearrowright$};
\node at (9,7) {$(T^\ast K_{\mathbb{C}})_{P,\mathrm{impl}} \tripleslash_{0} {\color{purple}{L_P^c}} = \overline{\mathcal{O}}_P$};
\draw (4.5,12)--(4.5,6);
\draw[->,thick] (0,9)--(0,8);
\node at (2,8.5) {Partial Implosion};
\node at (0,11.5) {\textsc{Partial HyperK\"ahler Implosion}};
\node at (8,11.5) {\textsc{Reduced Spaces}};
\end{tikzpicture}}
\caption{Top part: Universal HyperK\"ahler implosion. The symbol $\tripleslash_{0}$ denotes HyperK\"ahler quotient at $(0,0,0)$. $B$ is a Borel subgroup and $N$ is the maximal unipotent subgroup of $K_{\mathbb{C}}$. $\mathcal{N}$ is the nilpotent cone in $\mathfrak{k}_{\mathbb{C}}^\ast$. Bottom: Partial universal HyperK\"ahler implosion, where $\delta_P$ is the number of positive roots in the system defined by $P$. We use
$L_P^c$ to denote the compact form of the Levi.
When $P$ is minimal, i.e. $P=B$, the partial implosion reduces to the implosion.  }
\label{fig:implosionSummary2}
\end{figure}

Let us now recall the necessary mathematical preliminaries and review the main definitions of implosion spaces. 
The \emph{universal hyperK\"ahler implosion} \cite{dancer2013implosion} associated to a complex reductive group $G = K_{\mathbb C}$ is a space of complex dimension $\dim G + \mathrm{rank} \, G$ with an action of $G \times T_\mathbb C$ where $T_{\mathbb C}$ is the complexification of the maximal torus $T$ of $K$.
Moreover the hyperK\"ahler reductions by $T$, or equivalently the complex-symplectic reductions by $T_{\mathbb C}$, gives the Kostant varieties of $G$ -- in particular reduction at $0$ gives the nilpotent cone $\mathcal{N}$. The implosion also has a description as a nonreductive Geometric Invariant Theory (GIT) quotient by the maximal unipotent subgroup $N$ of $G$. 
This is summarized in the top part of Figure \ref{fig:implosionSummary2}. Note the analogy with (real) symplectic implosion, summarized in Figure \ref{fig:implosionSummary1}.

We can also consider {\em partial implosions}, which by analogy with the symplectic case \cite{Kirwan2011} we expect to be complex-symplectic quotients of $T^*G$ by the unipotent radical $U_P$ of a parabolic subgroup $P$. Explicitly, the partial implosion corresponding to $P$ should be
\begin{equation}
    (T^\ast G)_{P,\mathrm{impl}} =  (G \times {\mathfrak u}_P^\circ) \doubleslash U_P \, , 
\end{equation}
where ${\mathfrak u}_P^\circ$ denotes the annihilator of the Lie algebra of $U_P$, and $\doubleslash$ denotes the quotient in the GIT sense. The $G \times G$ action on $T^*G$ is now broken to an action of $G \times L_P$ where $L_P$ is the Levi subgroup of $P$ (recall that $P$ can be written as the semidirect product $U_P \rtimes L_P$ so $L_P$ normalizes $U_P$).  The explicit matrix descriptions of $P$, $U_P$, $\mathfrak{u}_P^{\circ}$ and $L_P$ are given for $\mathrm{GL}_4$ in Table \ref{tab:parabolicsA3}.  The 
complex dimension of the partial implosion is $2( \dim G - \dim U_P) = 2 \dim P$. The classical universal implosion of course corresponds to taking $P$ to be the Borel subgroup $B$, with $U_P$ equal to the maximal unipotent $N$ and the Levi being the complex maximal torus $T_{\mathbb{C}}$. This is summarized in the bottom part of Figure \ref{fig:implosionSummary2}. 

Assuming the conjecture that these partial implosions exist as algebraic varieties (i.e.\ that the ring of invariants for the quotient is finitely generated), we can make conjectures about their symplectic duals (the $\mathrm{SL}_n$ case was briefly discussed in \cite{DHK}).

One test of these conjectures is to consider the reduction of the partial implosion $(G \times {\mathfrak u}_P^\circ)\doubleslash U_P$ by the Levi group $L_P = P/U_P$. We should obtain the GIT quotient
\begin{equation} \label{quotient}
(G \times \mathfrak p^\circ ) \doubleslash P.
\end{equation}
Recall that if $P$ is the Borel $B$ we have the Springer resolution
\begin{equation}
T^*(G/B)= G \times_{B} \mathfrak b^\circ =
G \times_B \mathfrak n \rightarrow \mathcal N.
\end{equation}
Here we use an invariant inner product to identify
the annihilator $\mathfrak b^\circ$ with the
Lie algebra $\mathfrak n$ of the maximal unipotent
$N$. Moreover $\mathcal N$ denotes the nilpotent
cone, which is the closure of the regular nilpotent
orbit ${\mathcal O}_{\rm reg}$. The Springer resolution map is
\begin{equation}
(g,X) \mapsto gXg^{-1}.
\end{equation}
This is surjective, and is injective over the smooth locus ${\mathcal O}_{\rm reg}$ of $\mathcal N$. It is an affinisation map, in the 
sense that the base is the affine variety which is the Spec
of the ring of functions on the domain -- that is, 
$\mathcal N = (G \times \mathfrak b^\circ)\doubleslash B$, the quotient 
(\ref{quotient}).

\begin{table}[]
\makebox[\textwidth][c]{\scalebox{.9}{
\begin{tabular}{cccccccc}
\toprule 
Partition & Diagram & $P$ & $U_P$ & $\mathfrak{u}_P^{\circ}$ & $L_P$ & \begin{tabular}{c}
Partial Implosion \\ $ (G \times \mathfrak{u}_P^{\circ}) \doubleslash U_P  $
\end{tabular}  & \begin{tabular}{c}
Reduction  \\ by $L_P$
\end{tabular}   \\ \midrule 
\rowcolor{green!10} $[1^4]$ & \athree{white}{white}{white} & $\left( \begin{array}{cccc}
* & * & * & * \\
0 & * & * & * \\
0 & 0 & * & * \\
0 & 0 & 0 & * \\
\end{array} \right)$   & $\left( \begin{array}{cccc}
1 & * & * & * \\
0 & 1 & * & * \\
0 & 0 & 1 & * \\
0 & 0 & 0 & 1 \\
\end{array} \right)$ & $\left( \begin{array}{cccc}
* & 0 & 0 & 0 \\
* & * & 0 & 0 \\
* & * & * & 0 \\
* & * & * & * \\
\end{array} \right)$ &  $\left( \begin{array}{cccc}
* & 0 & 0 & 0 \\
0 & * & 0 & 0 \\
0 & 0 & * & 0 \\
0 & 0 & 0 & * \\
\end{array} \right)$ & $\mathrm{dim}_{\mathbb{H}} = 9$ & $\mathrm{dim}_{\mathbb{H}} = 6$\\ 
\rowcolor{blue!10} $[2,1^2]$ &  \begin{tabular}{c}
\athree{red}{white}{white} \\ \athree{white}{red}{white}  \\ \athree{white}{white}{red}
\end{tabular}  & $\left( \begin{array}{cccc}
* & * & * & * \\
* & * & * & * \\
0 & 0 & * & * \\
0 & 0 & 0 & * \\
\end{array} \right)$   & $\left( \begin{array}{cccc}
1 & 0 & * & * \\
0 & 1 & * & * \\
0 & 0 & 1 & * \\
0 & 0 & 0 & 1 \\
\end{array} \right)$ & $\left( \begin{array}{cccc}
* & * & 0 & 0 \\
* & * & 0 & 0 \\
* & * & * & 0 \\
* & * & * & * \\
\end{array} \right)$ & $\left( \begin{array}{cccc}
* & * & 0 & 0 \\
* & * & 0 & 0 \\
0 & 0 & * & 0 \\
0 & 0 & 0 & * \\
\end{array} \right)$ &  $\mathrm{dim}_{\mathbb{H}} = 10$ & $\mathrm{dim}_{\mathbb{H}} = 5$ \\ 
\rowcolor{red!10} $[2,2]$ &  \begin{tabular}{c}
\athree{red}{white}{red} 
\end{tabular}  & $\left( \begin{array}{cccc}
* & * & * & * \\
* & * & * & * \\
0 & 0 & * & * \\
0 & 0 & * & * \\
\end{array} \right)$   & $\left( \begin{array}{cccc}
1 & 0 & * & * \\
0 & 1 & * & * \\
0 & 0 & 1 & 0 \\
0 & 0 & 0 & 1 \\
\end{array} \right)$ & $\left( \begin{array}{cccc}
* & * & 0 & 0 \\
* & * & 0 & 0 \\
* & * & * & * \\
* & * & * & * \\
\end{array} \right)$ & $\left( \begin{array}{cccc}
* & * & 0 & 0 \\
* & * & 0 & 0 \\
0 & 0 & * & * \\
0 & 0 & * & * \\
\end{array} \right)$ &  $\mathrm{dim}_{\mathbb{H}} = 11$ & $\mathrm{dim}_{\mathbb{H}} = 4$ \\ 
\rowcolor{blue!10} $[3,1]$ & \begin{tabular}{c}
 \athree{red}{red}{white} \\ \athree{white}{red}{red}
\end{tabular}  & $\left( \begin{array}{cccc}
* & * & * & * \\
* & * & * & * \\
* & * & * & * \\
0 & 0 & 0 & * \\
\end{array} \right)$   & $\left( \begin{array}{cccc}
1 & 0 & 0 & * \\
0 & 1 & 0 & * \\
0 & 0 & 1 & * \\
0 & 0 & 0 & 1 \\
\end{array} \right)$ & $\left( \begin{array}{cccc}
* & * & * & 0 \\
* & * & * & 0 \\
* & * & * & 0 \\
* & * & * & * \\
\end{array} \right)$ & $\left( \begin{array}{cccc}
* & * & * & 0 \\
* & * & * & 0 \\
* & * & * & 0 \\
0 & 0 & 0 & * \\
\end{array} \right)$ & $\mathrm{dim}_{\mathbb{H}} = 12$ & $\mathrm{dim}_{\mathbb{H}} = 3$ \\
\rowcolor{blue!10} $[4]$ & \begin{tabular}{c}
\athree{red}{red}{red} 
\end{tabular}  & $\left( \begin{array}{cccc}
* & * & * & * \\
* & * & * & * \\
* & * & * & * \\
* & * & * & * \\
\end{array} \right)$   & $\left( \begin{array}{cccc}
1 & 0 & 0 & 0 \\
0 & 1 & 0 & 0 \\
0 & 0 & 1 & 0 \\
0 & 0 & 0 & 1 \\
\end{array} \right)$ & $\left( \begin{array}{cccc}
* & * & * & * \\
* & * & * & * \\
* & * & * & * \\
* & * & * & * \\
\end{array} \right)$ & $\left( \begin{array}{cccc}
* & * & * & * \\
* & * & * & * \\
* & * & * & * \\
* & * & * & * \\
\end{array} \right)$ & $\mathrm{dim}_{\mathbb{H}} = 15$ & $\mathrm{dim}_{\mathbb{H}} = 0$ \\ \bottomrule
\end{tabular}}}
    \caption{List of parabolic subgroups of $\mathrm{GL}_4$. Each parabolic is labeled by an ordered partition, or equivalently a subset of the simple roots, as shown in the second column. The parabolic subgroup, unipotent radical and Levi subgroup are shown. We then give the quaternionic dimensions of the partial implosions and their GIT reduction by the Levi. The line shaded in green is the usual implosion. The blue lines are the hooks. Note that the
    partition given here is dual to that of the corresponding Jordan type.}
    \label{tab:parabolicsA3}
\end{table}

For a general parabolic we have an analogous picture, the
{\em partial Springer resolution}
\begin{equation}
G \times_{P} \mathfrak p^\circ = G \times_{P} {\mathfrak u}_P \rightarrow \overline{{\mathcal O}_P}.
\end{equation}
The domain can be viewed as $T^*(G/P)$, the cotangent bundle of the
variety $\mathcal P$ of parabolics conjugate to $P$.

The orbit ${\mathcal O}_P$ is the Richardson orbit associated to the
parabolic $P$, which is characterised by the fact that its intersection
with the nilradical ${\mathfrak u}_{P}$ is dense in ${\mathfrak u}_P$. It is possible for two
non-conjugate parabolics to give the same Richardson
orbit. (See \cite{CMcG} for general background on nilpotent orbits).

In the $\mathrm{SL}_n$ case all orbits are Richardson and the partial Springer
resolution is injective over $\mathcal O_{\rm reg}$.

For other groups the partial Springer map may be finite-to-one
rather than injective on the smooth locus.
Moreover, not all nilpotent orbits are Richardson (in particular
the minimal orbit is not Richardson except for $\mathrm{SL}_n$). We expect
that reduction by the Levi should still give the closure of the
Richardson  orbit if this is normal and the partial Springer map
is injective over the  smooth locus (these
properties are automatic in the $\mathrm{SL}_n$ case).

\section{Unitary partial implosion}
\label{sec:unitary}

In this section, we provide magnetic quivers for the partial implosions for $G=\mathrm{SL}_n$. 
In this case a choice of parabolic $P$ corresponds to an ordered partition 
\begin{equation}
    P \quad \leftrightarrow \quad n = n_1 + \ldots + n_r \qquad \textrm{with} \qquad  n_1 \geq \cdots \geq n_r > 0 \, , 
\end{equation}
and the associated Levi is then
\begin{equation} \label{Levi}
L_P = \mathrm{S}(\mathrm{GL}_{n_{1}} \times \ldots \times \mathrm{GL}_{n_r}).
\end{equation}
We want therefore a space which has hyperK\"ahler symmetry
\begin{equation}
\label{globalSym}
\mathrm{SU}(n) \times \mathrm{S}(\mathrm{U}(n_1) \times \ldots \times \mathrm{U}(n_r)).
\end{equation}
We consider the diagram obtained by taking the basic $A_{n}$ quiver for the nilpotent cone,
removing the top $\mathrm{SU}(n)$ flavour node, and then attaching to the $n-1$
dimensional node $r$ 
legs, each of them an $A_{n_i}$ quiver with the dimension $n_i$ node
next to the $n-1$ node:  
\begin{equation}
\label{implodedQuiver}
(T^\ast \mathrm{SL}_n)_{P,\mathrm{impl}} =  \mathcal{C}\left( \raisebox{-.5\height}{\begin{tikzpicture}
	\tikzset{node distance = .8cm};
	\node (g1) at (1,0) [gauge,label=below:1] {};
	\node (g2) at (2,0) {$\cdots$};
	\node (g3) at (3,0) [gauge,label=below:$n-1$] {};
	\node (g41) at (5,1) [gaugeb,label=below:$n_1$] {};
	\node (g42) at (6,1) [gauge,label=below:$n_1-1$] {};
	\node (g43) at (7,1) {$\cdots$};
	\node (g44) at (8,1) [gauge,label=below:$1$] {};
	\node (g61) at (5,0) {$\vdots$};
	\node (g62) at (6,0) {$\vdots$};
	\node (g64) at (8,0) {$\vdots$};
	\node (g71) at (5,-1) [gaugeb,label=below:$n_r$] {};
	\node (g72) at (6,-1) [gauge,label=below:$n_r-1$] {};
	\node (g73) at (7,-1) {$\cdots$};
	\node (g74) at (8,-1) [gauge,label=below:$1$] {};
	\draw (g1)--(g2)--(g3);
	\draw (g3)--(g41)--(g42)--(g43)--(g44);
	\draw (g3)--(g71)--(g72)--(g73)--(g74);
	\end{tikzpicture}} \right)
\end{equation}
The $n-1$ node therefore remains balanced as $n = \sum_{i=1}^{r} n_i$. Moreover all the nodes in the legs are balanced except for the $n_i$ nodes: the non-balanced nodes are depicted in black in the quiver above. 
Notice that the balance at the non-balanced $n_i$ nodes is $n - n_i-2$ so these nodes
are 
\begin{itemize}
    \item Bad for the trivial partition $n$, when the parabolic is $G$ itself and no implosion takes place. We shall not consider this case further. 
    \item Minimally unbalanced if the partition is $(n-1) +1$, with $n_i=n-1$. In that case the Coulomb branch of (\ref{implodedQuiver}) reduces to the flat space $\mathbb{H}^{n(n-1)}$. 
    \item Balanced if the partition is $(n-2) + 2$ (respectively $(n-2)+1+1$),  with $n_i =n-2$. In this case the symmetry (\ref{globalSym}), $\mathrm{SU}(n) \times \mathrm{SU}(n-2) \times \mathrm{SU}(2) \times \mathrm{U}(1)$ (resp. $\mathrm{SU}(n) \times \mathrm{SU}(n-2) \times \mathrm{U}(1)^2$) enhances to $\mathrm{SU}(2n-2)  \times \mathrm{SU}(2)$ (resp. $\mathrm{SU}(2n-2)  \times \mathrm{U}(1)$). 
    \item  Of positive balance in all other cases, and the symmetry is (\ref{globalSym}). 
\end{itemize}

In \cite{DHK} the quaternionic dimension of the partial implosion was shown to be 
\begin{equation}
\mathrm{dim}_{\mathbb{H}} (T^\ast \mathrm{SL}_n)_{P,\mathrm{impl}} = \frac{1}{2} \left(n^2 - 2 + \sum_{i=1}^{r} n_i^2 \right),
\end{equation}
which equals the rank of the gauge group in the magnetic quiver in (\ref{implodedQuiver}), 
\begin{equation}
\mathrm{S} \left( \mathrm{U}(1) \times \ldots \times \mathrm{U}(n-1) \times \prod_{i=1}^{r} \mathrm{U}(1) \times \ldots \times \mathrm{U}(n_i) \right)
\end{equation}
in accordance with Nakajima's equality \cite{Nakajima:2018}. The Levi $L_P$ has complex dimension
\begin{equation}
    \mathrm{dim}_{\mathbb{C}} L_P = -1 + \sum_{i=1}^{r} n_i^2 \, , 
\end{equation}
so the quaternionic dimension of the reduction is 
 \begin{eqnarray}
 \label{equationReduction}
   \mathrm{dim}_{\mathbb{H}}\left[ (T^\ast \mathrm{SL}_n)_{P,\mathrm{impl}} \tripleslash_{0} {L_P^c}  \right] &=&  \frac{1}{2} \left(n^2 - 2 + \sum_{i=1}^{r} n_i^2 \right) -  \left(\sum_{i=1}^{r} n_i^2  -1 \right) \nonumber \\ &=& \frac{1}{2} \left(n^2  -  \sum_{i=1}^{r} n_i^2 \right) \, . 
\end{eqnarray}
The Richardson orbit $\mathcal O_{P}$ corresponding to the
parabolic has Jordan block sizes given by the \emph{dual} partition to $(n_1,\ldots,n_r)$.
This has complex dimension $n^2 -\sum_{i=1}^{r} n_i^2$, agreeing with (\ref{equationReduction}). 

Of course we recover the classical
implosion by taking the partition
$n=1+ \dots +1$, and this situation
has been explored in detail in \cite{DHK}.
The quiver graph in this case is an example of that in the
splaying (or 0-fission) construction introduced by Boalch in \cite{Boalch2008}, but note that in our case the bouquet nodes are gauge rather than flavour nodes (closed rather than open in the notation of \cite{Boalch2008}).

\medskip
The next step, now that we have checked that the dimensions work out, is to attempt to compute reductions of the Coulomb branch by the Levi. As we are now quotienting by a non-Abelian group, this is a much more difficult problem than in the original implosion situation studied in \cite{DHK}.

In the rest of this section, we perform checks of the main claim (\ref{implodedQuiver}) for a class of partitions $n = k + 1 + \dots + 1$ dubbed \emph{hooks}, because of the shape of the corresponding Young diagram. These partitions allow us to verify through an explicit computation the equality between the HyperK\"ahler reduced partial implosion and the corresponding nilpotent orbit closure. 
We first consider the particular case $k=2$, before turning to the general case.

\subsection{Subregular case}

In this section we consider the {\em subregular} case, which is closest to the standard implosion. We now take 
\begin{equation}
\label{subregularPartition}
P  \quad \leftrightarrow \quad n = 2 + 1 + \ldots + 1
\end{equation}
as our partition. So we have one $A_2$ leg in our quiver diagram, together with $n-2$ one-dimensional nodes, see the left part of equation (\ref{implosionSubregul}). 
In the $n=4$ case, uniquely, the 2-dimensional
node in the $A_2$ leg is also balanced so we have a
symmetry enhancement to $\mathrm{U}(6)$.
The (compact form of the) Levi group by which we can hyperK\"ahler reduce is
\begin{equation}
    L_P^c = \mathrm{S}(\mathrm{U}(2) \times \mathrm{U}(1)^{n-2}) \cong \mathrm{SU}(2) \times \mathrm{U}(1)^{n-2} \, .
\end{equation}

We expect to obtain the closure of the subregular orbit, whose Jordan type $[n-1,1]$ is dual to the above partition (\ref{subregularPartition}). The complex dimension of this orbit is $n^2- n -2$.  In terms of quivers, the hyperK\"ahler quotient takes the following form:
\begin{equation}
\label{implosionSubregul}
\raisebox{-.5\height}{\scalebox{.8}{
\begin{tikzpicture}
	\begin{pgfonlayer}{nodelayer}
		\node [style=gauge3] (0) at (0, 0) {};
		\node [style=gauge3] (3) at (0.5, 1.25) {};
		\node [style=gauge3] (5) at (1.25, 0.25) {};
		\node [style=gauge3] (7) at (0.625, -1.2) {};
		\node [style=gauge3] (8) at (1, 2.5) {};
		\node [style=none] (9) at (0.5, 2.5) {1};
		\node [style=none] (10) at (0, 1.25) {2};
		\node [style=none] (11) at (-0.5, -0.5) {$n{-}1$};
		\node [style=none] (15) at (0.625, -1.7) {1};
		\node [style=none] (17) at (1.75, 0.25) {1};
		\node [style=none] (18) at (-0.5, 0) {};
		\node [style=none] (19) at (-1.5, 0) {};
		\node [style=gauge3] (20) at (-2, 0) {};
		\node [style=gauge3] (21) at (-3, 0) {};
		\node [style=gauge3] (22) at (-4, 0) {};
		\node [style=none] (23) at (-4, -0.5) {1};
		\node [style=none] (24) at (-3, -0.5) {2};
		\node [style=none] (25) at (-2, -0.5) {3};
		\node [style=bd] (26) at (1.275, -0.275) {};
		\node [style=bd] (27) at (0.975, -0.775) {};
		\node [style=none] (28) at (0.875, -1.825) {};
		\node [style=none] (29) at (1.875, 0.175) {};
		\node [style=none] (30) at (2, -1.075) {$n{-}2$};
		\node [style=gauge3] (31) at (12, 0) {};
		\node [style=none] (38) at (12, -0.5) {$n{-}2$};
		\node [style=none] (42) at (8.5, 0) {};
		\node [style=gauge3] (43) at (8, 0) {};
		\node [style=gauge3] (44) at (7, 0) {};
		\node [style=gauge3] (45) at (6, 0) {};
		\node [style=none] (46) at (6, -0.5) {1};
		\node [style=none] (47) at (7, -0.5) {2};
		\node [style=none] (48) at (8, -0.5) {3};
		\node [style=none] (54) at (3, 0) {};
		\node [style=none] (55) at (5, 0) {};
		\node [style=gauge3] (56) at (11, 0) {};
		\node [style=flavour2] (57) at (12, 1) {};
		\node [style=none] (58) at (12, 1.5) {$n{-}2$};
		\node [style=none] (60) at (11, -0.5) {$n{-}2$};
		\node [style=flavour2] (61) at (11, 1) {};
		\node [style=none] (62) at (11, 1.5) {$1$};
		\node [style=gauge3] (63) at (10, 0) {};
		\node [style=none] (64) at (10, -0.5) {$n{-}3$};
		\node [style=none] (65) at (9.5, 0) {};
		\node [style=none] (66) at (4.75, 1) {\parbox{5cm}{$\mathrm{SU}(2)\times \mathrm{U}(1)^{n-2}$ \\hyperK\"ahler Quotient}};
		\node [style=none] (67) at (-1, 0) {$\dots$};
		\node [style=none] (68) at (9, 0) {$\dots$};
	\end{pgfonlayer}
	\begin{pgfonlayer}{edgelayer}
		\draw (0) to (3);
		\draw (0) to (5);
		\draw (0) to (7);
		\draw (3) to (8);
		\draw (20) to (19.center);
		\draw (22) to (21);
		\draw (21) to (20);
		\draw (18.center) to (0);
		\draw [style=brace] (29.center) to (28.center);
		\draw (43) to (42.center);
		\draw (45) to (44);
		\draw (44) to (43);
		\draw [style=->] (54.center) to (55.center);
		\draw (57) to (31);
		\draw (61) to (56);
		\draw (56) to (31);
		\draw (56) to (63);
		\draw (65.center) to (63);
	\end{pgfonlayer}
\end{tikzpicture}}}
\end{equation}
The quiver on the right-hand side has Coulomb branch equal to the closure of the subregular nilpotent orbit of $\mathfrak{sl}_n$, as can be checked by quiver subtraction from the $T[\mathrm{SU}(n)]$ quiver. In order to compute the hyperK\"ahler quotient in (\ref{implosionSubregul}), we use the fact that performing a hyperK\"ahler quotient by $H$ on a quiver is equivalent to gauging a subgroup $H$ of the global symmetry of the 3d mirror, as reviewed in Section \ref{sec:phys}. 

As a check of this procedure, we look at an explicit computation of the Coulomb branch Hilbert series with $n=5$ with Jordan type $[4,1]$. The quiver is drawn with the following fugacities:
\begin{equation}
  \raisebox{-.5\height}{\begin{tikzpicture}
	\begin{pgfonlayer}{nodelayer}
		\node [style=gauge3] (0) at (0, 0) {};
		\node [style=gauge3] (3) at (0.5, 1.25) {};
		\node [style=gauge3] (5) at (1.25, 0.5) {};
		\node [style=gauge3] (8) at (1, 2.5) {};
		\node [style=none] (9) at (0.5, 2.5) {1};
		\node [style=none] (10) at (0, 1.25) {2};
		\node [style=none] (15) at (0.975, -1.7) {1};
		\node [style=none] (17) at (1.75, 0.5) {1};
		\node [style=gauge3] (20) at (-1, 0) {};
		\node [style=gauge3] (21) at (-2, 0) {};
		\node [style=gauge3] (22) at (-3, 0) {};
		\node [style=none] (23) at (-3, -0.5) {1};
		\node [style=none] (24) at (-2, -0.5) {2};
		\node [style=none] (25) at (-1, -0.5) {3};
		\node [style=none] (26) at (0, -0.5) {4};
		\node [style=gauge3] (27) at (1.25, -0.5) {};
		\node [style=none] (28) at (2.25, -1) { $\color{red}{z_2}$};
		\node [style=flavour2] (29) at (1, -1.25) {};
		\node [style=none] (30) at (2.25, 0.5) { $\color{red}{z_1}$};
		\node [style=none] (31) at (1, 3) { $\color{red}{x}$};
		\node [style=none] (32) at (1, 1.25) { $\color{red}{q}$};
		\node [style=none] (33) at (1.75, -0.75) {1};
	\end{pgfonlayer}
	\begin{pgfonlayer}{edgelayer}
		\draw (0) to (3);
		\draw (0) to (5);
		\draw (3) to (8);
		\draw (22) to (21);
		\draw (21) to (20);
		\draw (20) to (0);
		\draw (0) to (27);
		\draw (0) to (29);
	\end{pgfonlayer}
\end{tikzpicture}

}
\label{beforequotient}
\end{equation}
where $z_1,z_2,q$ are the fugacities of the three $\mathrm{U}(1)$ subgroups and $x$ is the fugacity of the $\mathrm{SU}(2)$ subgroup. The Coulomb branch Hilbert series is $\widetilde{\mathrm{HS}}_{[4,1]}(z_1,z_2,q,x;t) $. Note that one node is ungauged to fix an overall $\mathrm{U}(1)$ which is trivially acting. The hyperK\"ahler quotient takes $\widetilde{\mathrm{HS}}_{[4,1]}(z_1,z_2,q,x;t) $ to $\mathrm{HS}_{[4,1]}(t)$, in accordance with our conjecture:
\begin{equation}
\label{HS41}
\begin{split}
   \oint \frac{ \mathrm{d} z_1  \, \mathrm{d} z_2 \, \mathrm{d} q \, \mathrm{d} x}{(2\pi i)^4 z_1z_2q}   &\frac{1-x^2}{x} (1-t^2)^4(1-x^2t^2) \left(1-\frac{t^2}{x^2} \right)\widetilde{\mathrm{HS}}_{[4,1]}(z_1,z_2,q,x;t) \\=  & 1+24t^2+299t^4\dots = \mathrm{HS}_{[4,1]}(t),
      \end{split}
\end{equation}
where
\begin{equation}
\begin{split}
  \widetilde{\mathrm{HS}}_{[4,1]}(z_1,z_2,q,x;t) & =  1+\left(28+\frac{1}{x^2}+x^2\right )t^2 +  \left(406+\frac{1}{x^4}+ \frac{28}{x^2}+28x^2+x^4 \right. \\ & \left. +\frac{5}{z_1}+ 5z_1+ \frac{5}{z_2}+5z_2+\frac{5}{q^2z_1z_2}+5q^2z_1z_2\right)t^4 +\dots
   \end{split}
\end{equation}
Note that the term $24t^2$ in (\ref{HS41}) indicates the expected dimension of the global symmetry group. The Hilbert series (\ref{HS41}) indeed agrees with that of the expected nilpotent orbit closure \cite{Hanany:2016gbz}.

\subsection{Hooks}

We next generalise this to consider {\em hook partitions}
\begin{equation}
P  \quad \leftrightarrow \quad  n = k + 1 + \ldots + 1 \, , 
\end{equation}
for which the quiver is 
\begin{equation}
\raisebox{-.5\height}{\begin{tikzpicture}
	\begin{pgfonlayer}{nodelayer}
		\node [style=gauge3] (0) at (0, 0) {};
		\node [style=gauge3] (1) at (1.15, 0) {};
		\node [style=gauge3] (2) at (0.75, 1.025) {};
		\node [style=gauge3] (3) at (-0.625, 1.05) {};
		\node [style=gauge3] (4) at (2.275, 0) {};
		\node [style=none] (5) at (2.275, -0.5) {$k{-}1$};
		\node [style=none] (6) at (1.15, -0.5) {$k$};
		\node [style=none] (7) at (0, -0.5) {$n{-}1$};
		\node [style=none] (8) at (-1.125, 1.05) {1};
		\node [style=none] (9) at (1.25, 1) {1};
		\node [style=none] (10) at (-0.5, 0) {};
		\node [style=none] (11) at (-1.5, 0) {};
		\node [style=gauge3] (12) at (-2, 0) {};
		\node [style=gauge3] (13) at (-3, 0) {};
		\node [style=gauge3] (14) at (-4, 0) {};
		\node [style=none] (15) at (-4, -0.5) {1};
		\node [style=none] (16) at (-3, -0.5) {2};
		\node [style=none] (17) at (-2, -0.5) {3};
		\node [style=none] (20) at (0.775, 1.325) {};
		\node [style=none] (21) at (-0.725, 1.325) {};
		\node [style=none] (22) at (0, 1.925) {$n{-}k$};
		\node [style=none] (44) at (3.25, 0.025) {\dots};
		\node [style=none] (45) at (4.25, -0.475) {1};
		\node [style=gauge3] (46) at (4.25, 0.025) {};
		\node [style=none] (47) at (0.05, 1) {\dots};
		\node [style=none] (48) at (2.75, 0) {};
		\node [style=none] (49) at (3.75, 0.025) {};
		\node [style=none] (50) at (-1, 0) {$\dots$};
	\end{pgfonlayer}
	\begin{pgfonlayer}{edgelayer}
		\draw (0) to (1);
		\draw (0) to (2);
		\draw (0) to (3);
		\draw (1) to (4);
		\draw (12) to (11.center);
		\draw (14) to (13);
		\draw (13) to (12);
		\draw (10.center) to (0);
		\draw [style=brace] (21.center) to (20.center);
		\draw (4) to (48.center);
		\draw (49.center) to (46);
	\end{pgfonlayer}
\end{tikzpicture}}
\label{hookquiver}
\end{equation}
where the Coulomb branch global symmetry is $\mathrm{SU}(n)\times  \mathrm{SU}(k) \times \mathrm{U}(1)^{n-k}$. The Levi is now 
\begin{equation}
\label{leviHook}
    L_P^c = \mathrm{SU}(k) \times \mathrm{U}(1)^{n-k} \, , 
\end{equation}
and on hyperK\"ahler reduction we expect to
obtain the closure of the $[n-k+1, 1^{k-1}] $ orbit, whose complex dimension is $n^2 - n - (k^2 -k)$. 

Taking the hyperK\"ahler quotients over the $k$ leg unfortunately runs into problems of incomplete Higgsing. However, the advantage of the hook quivers is that they have known 3d mirrors. The Coulomb branch of \eqref{hookquiver} is the Higgs branch of:
\begin{equation}
\raisebox{-.5\height}{\begin{tikzpicture}
	\begin{pgfonlayer}{nodelayer}
		\node [style=gauge3] (0) at (-1.75, 0) {};
		\node [style=gauge3] (1) at (-0.25, 0) {};
		\node [style=gauge3] (2) at (2.25, 0) {};
		\node [style=none] (4) at (-2, -0.5) {$\mathrm{SU}(n{-}1)$};
		\node [style=none] (5) at (-0.25, -0.5) {$\mathrm{SU}(n{-}2)$};
		\node [style=none] (6) at (1, 0) {\dots};
		\node [style=none] (7) at (0.25, 0) {};
		\node [style=none] (8) at (1.75, 0) {};
		\node [style=none] (9) at (2.25, -0.5) {$\mathrm{SU}(k{+}1)$};
		\node [style=none] (13) at (-1.75, 1.5) {$n$};
		\node [style=none] (15) at (2.25, 1.5) {$k$};
		\node [style=flavour2] (16) at (-1.75, 1) {};
		\node [style=flavour2] (18) at (2.25, 1) {};
	\end{pgfonlayer}
	\begin{pgfonlayer}{edgelayer}
		\draw (1) to (7.center);
		\draw (8.center) to (2);
		\draw (0) to (1);
		\draw (16) to (0);
		\draw (18) to (2);
	\end{pgfonlayer}
\end{tikzpicture}}
\label{hookmirror}
\end{equation}

\begin{figure}
    \centering
\scalebox{0.75}{
\begin{tikzpicture}
	\begin{pgfonlayer}{nodelayer}
		\node [style=gauge3] (0) at (-0.75, 0) {};
		\node [style=gauge3] (1) at (0.5, 0) {};
		\node [style=gauge3] (2) at (3, 0) {};
		\node [style=none] (4) at (-1, -0.5) {$\mathrm{SU}(n{-}1)$};
		\node [style=none] (5) at (0.5, 0.5) {$\mathrm{SU}(n{-}2)$};
		\node [style=none] (6) at (1.75, 0) {\dots};
		\node [style=none] (7) at (1, 0) {};
		\node [style=none] (8) at (2.5, 0) {};
		\node [style=none] (9) at (3, -0.5) {$\mathrm{SU}(k{+}1)$};
		\node [style=none] (13) at (-0.75, 1.5) {$n$};
		\node [style=none] (15) at (3, 1.5) {$k$};
		\node [style=flavour2] (16) at (-0.75, 1) {};
		\node [style=flavour2] (18) at (3, 1) {};
		\node [style=gauge3] (19) at (-9.25, 0) {};
		\node [style=gauge3] (20) at (-8.1, 0) {};
		\node [style=gauge3] (21) at (-8.5, 1.025) {};
		\node [style=gauge3] (22) at (-9.875, 1.05) {};
		\node [style=gauge3] (23) at (-6.975, 0) {};
		\node [style=none] (24) at (-6.975, -0.5) {$k{-}1$};
		\node [style=none] (25) at (-8.1, -0.5) {$k$};
		\node [style=none] (26) at (-9.25, -0.5) {$n{-}1$};
		\node [style=none] (27) at (-10.375, 1.05) {1};
		\node [style=none] (28) at (-8, 1) {1};
		\node [style=none] (29) at (-9.75, 0) {};
		\node [style=none] (30) at (-10.75, 0) {};
		\node [style=gauge3] (31) at (-11.25, 0) {};
		\node [style=gauge3] (32) at (-12.25, 0) {};
		\node [style=gauge3] (33) at (-13.25, 0) {};
		\node [style=none] (34) at (-13.25, -0.5) {1};
		\node [style=none] (35) at (-12.25, -0.5) {2};
		\node [style=none] (36) at (-11.25, -0.5) {3};
		\node [style=none] (37) at (-8.475, 1.325) {};
		\node [style=none] (38) at (-9.975, 1.325) {};
		\node [style=none] (39) at (-9.25, 1.925) {$n{-}k$};
		\node [style=none] (40) at (-6, 0) {\dots};
		\node [style=none] (41) at (-5, -0.5) {1};
		\node [style=gauge3] (42) at (-5, 0) {};
		\node [style=none] (43) at (-9.2, 1) {\dots};
		\node [style=none] (44) at (-6.5, 0) {};
		\node [style=none] (45) at (-5.5, 0) {};
		\node [style=none] (46) at (-4, 0) {};
		\node [style=none] (47) at (-2, 0) {};
		\node [style=gauge3] (48) at (-0.75, -6.5) {};
		\node [style=gauge3] (49) at (0.5, -6.5) {};
		\node [style=gauge3] (50) at (3, -6.5) {};
		\node [style=none] (51) at (-1, -7) {$\mathrm{U}(n{-}1)$};
		\node [style=none] (52) at (0.5, -6) {$\mathrm{U}(n{-}2)$};
		\node [style=none] (53) at (1.75, -6.5) {\dots};
		\node [style=none] (54) at (1, -6.5) {};
		\node [style=none] (55) at (2.5, -6.5) {};
		\node [style=none] (57) at (-0.75, -5) {$n$};
		\node [style=none] (58) at (3, -5) {$k$};
		\node [style=flavour2] (59) at (-0.75, -5.5) {};
		\node [style=flavour2] (60) at (3, -5.5) {};
		\node [style=gauge3] (61) at (-9.25, -6.5) {};
		\node [style=gauge3] (62) at (-8.1, -6.5) {};
		\node [style=gauge3] (65) at (-6.975, -6.5) {};
		\node [style=none] (66) at (-6.975, -7) {$k{-}1$};
		\node [style=none] (67) at (-8.1, -7) {$k$};
		\node [style=none] (68) at (-9.25, -7) {$n{-}1$};
		\node [style=none] (71) at (-9.75, -6.5) {};
		\node [style=none] (72) at (-10.75, -6.5) {};
		\node [style=gauge3] (73) at (-11.25, -6.5) {};
		\node [style=gauge3] (74) at (-12.25, -6.5) {};
		\node [style=gauge3] (75) at (-13.25, -6.5) {};
		\node [style=none] (76) at (-13.25, -7) {1};
		\node [style=none] (77) at (-12.25, -7) {2};
		\node [style=none] (78) at (-11.25, -7) {3};
		\node [style=none] (82) at (-6, -6.5) {\dots};
		\node [style=none] (83) at (-5, -7) {1};
		\node [style=gauge3] (84) at (-5, -6.5) {};
		\node [style=none] (86) at (-6.5, -6.5) {};
		\node [style=none] (87) at (-5.5, -6.5) {};
		\node [style=none] (88) at (-4, -6.5) {};
		\node [style=none] (89) at (-2, -6.5) {};
		\node [style=none] (90) at (-9.25, -1.75) {};
		\node [style=none] (91) at (-9.25, -3.5) {};
		\node [style=none] (92) at (1, -1.75) {};
		\node [style=none] (93) at (1, -3.5) {};
		\node [style=none] (94) at (-12.25, -2.75) {hyperK\"ahler quotient $\mathrm{U}(1)^{n-k-1}$};
		\node [style=none] (96) at (2.75, -2.75) {Gauging $\mathrm{U}(1)^{n-k-1}$};
		\node [style=flavour2] (97) at (-9.25, -5.5) {};
		\node [style=none] (98) at (-9.25, -5) {$n{-}k$};
		\node [style=gauge3] (99) at (-0.75, -12.5) {};
		\node [style=gauge3] (100) at (0.5, -12.5) {};
		\node [style=gauge3] (101) at (3, -12.5) {};
		\node [style=none] (102) at (-1, -13) {$\mathrm{U}(n{-}1)$};
		\node [style=none] (103) at (0.5, -12) {$\mathrm{U}(n{-}2)$};
		\node [style=none] (104) at (1.75, -12.5) {\dots};
		\node [style=none] (105) at (1, -12.5) {};
		\node [style=none] (106) at (2.5, -12.5) {};
		\node [style=none] (107) at (3, -13) {$\mathrm{U}(k{+}1)$};
		\node [style=none] (108) at (-0.75, -11) {$n$};
		\node [style=none] (109) at (3, -11) {$\mathrm{U}(k)$};
		\node [style=flavour2] (110) at (-0.75, -11.5) {};
		\node [style=gauge3] (112) at (-6.5, -12.25) {};
		\node [style=none] (117) at (-6.5, -12.75) {$k$};
		\node [style=none] (119) at (-10.75, -12.25) {};
		\node [style=gauge3] (120) at (-11.25, -12.25) {};
		\node [style=gauge3] (121) at (-12.25, -12.25) {};
		\node [style=gauge3] (122) at (-13.25, -12.25) {};
		\node [style=none] (123) at (-13.25, -12.75) {1};
		\node [style=none] (124) at (-12.25, -12.75) {2};
		\node [style=none] (125) at (-11.25, -12.75) {3};
		\node [style=none] (131) at (-4, -12.25) {};
		\node [style=none] (132) at (-2, -12.25) {};
		\node [style=none] (133) at (-9.25, -8) {};
		\node [style=none] (134) at (-9.25, -9.75) {};
		\node [style=none] (135) at (1, -8) {};
		\node [style=none] (136) at (1, -9.75) {};
		\node [style=none] (137) at (-12.25, -9) {hyperK\"ahler quotient $\mathrm{U}(k)$};
		\node [style=flavour2] (139) at (-6.5, -11.25) {};
		\node [style=none] (140) at (-6.5, -10.75) {$k$};
		\node [style=none] (141) at (2.75, -9) {Gauging $\mathrm{U}(k)$};
		\node [style=gauge3] (142) at (3, -11.5) {};
		\node [style=gauge3] (143) at (-7.25, -12.25) {};
		\node [style=none] (144) at (-7.25, -12.75) {$k$};
		\node [style=none] (145) at (-10.25, -12.25) {\dots};
		\node [style=none] (146) at (-9.75, -12.25) {};
		\node [style=gauge3] (147) at (-9.25, -12.25) {};
		\node [style=none] (148) at (-9.25, -12.75) {$k$};
		\node [style=none] (149) at (-8.25, -12.25) {\dots};
		\node [style=none] (150) at (-7.75, -12.25) {};
		\node [style=none] (151) at (-8.75, -12.25) {};
		\node [style=flavour2] (152) at (-9.25, -11.25) {};
		\node [style=none] (153) at (-9.25, -10.75) {1};
		\node [style=none] (154) at (-9.25, -13.25) {};
		\node [style=none] (155) at (-6.5, -13.25) {};
		\node [style=none] (156) at (-7.9, -13.75) {$n{-}k$};
		\node [style=none] (157) at (-3, 0.5) {3d mirror};
		\node [style=none] (158) at (-3, -6) {3d mirror};
		\node [style=none] (159) at (-3, -11.75) {3d mirror};
		\node [style=none] (160) at (-10.25, 0) {\dots};
		\node [style=none] (161) at (-10.25, -6.5) {\dots};
		\node [style=none] (162) at (1, -13.75) {};
		\node [style=none] (163) at (1, -16) {};
		\node [style=gauge3] (164) at (-0.75, -17.75) {};
		\node [style=gauge3] (165) at (0.5, -17.75) {};
		\node [style=gauge3] (166) at (3, -17.75) {};
		\node [style=none] (167) at (-0.75, -18.25) {$\mathrm{U}(n{-}k)$};
		\node [style=none] (168) at (0.5, -17.25) {$\mathrm{U}(n{-}k{-}1)$};
		\node [style=none] (169) at (1.75, -17.75) {\dots};
		\node [style=none] (170) at (1, -17.75) {};
		\node [style=none] (171) at (2.5, -17.75) {};
		\node [style=none] (172) at (3, -18.25) {$\mathrm{U}(2)$};
		\node [style=none] (173) at (-0.75, -16.25) {$n$};
		\node [style=none] (174) at (3, -16.25) {$\mathrm{U}(1)$};
		\node [style=flavour2] (175) at (-0.75, -16.75) {};
		\node [style=gauge3] (176) at (3, -16.75) {};
		\node [style=none] (177) at (4, -15) {\parbox{5cm}{Same Higgs branch\\ when $FI \neq 0$}};
		\node [style=none] (178) at (3, -7) {$\mathrm{U}(k{+}1)$};
	\end{pgfonlayer}
	\begin{pgfonlayer}{edgelayer}
		\draw (1) to (7.center);
		\draw (8.center) to (2);
		\draw (0) to (1);
		\draw (16) to (0);
		\draw (18) to (2);
		\draw (19) to (20);
		\draw (19) to (21);
		\draw (19) to (22);
		\draw (20) to (23);
		\draw (31) to (30.center);
		\draw (33) to (32);
		\draw (32) to (31);
		\draw (29.center) to (19);
		\draw [style=brace] (38.center) to (37.center);
		\draw (23) to (44.center);
		\draw (45.center) to (42);
		\draw [style=->] (46.center) to (47.center);
		\draw [style=->] (47.center) to (46.center);
		\draw (49) to (54.center);
		\draw (55.center) to (50);
		\draw (48) to (49);
		\draw (59) to (48);
		\draw (60) to (50);
		\draw (61) to (62);
		\draw (62) to (65);
		\draw (73) to (72.center);
		\draw (75) to (74);
		\draw (74) to (73);
		\draw (71.center) to (61);
		\draw (65) to (86.center);
		\draw (87.center) to (84);
		\draw [style=->] (88.center) to (89.center);
		\draw [style=->] (89.center) to (88.center);
		\draw [style=->] (90.center) to (91.center);
		\draw [style=->] (92.center) to (93.center);
		\draw (97) to (61);
		\draw (100) to (105.center);
		\draw (106.center) to (101);
		\draw (99) to (100);
		\draw (110) to (99);
		\draw (120) to (119.center);
		\draw (122) to (121);
		\draw (121) to (120);
		\draw [style=->] (131.center) to (132.center);
		\draw [style=->] (132.center) to (131.center);
		\draw [style=->] (133.center) to (134.center);
		\draw [style=->] (135.center) to (136.center);
		\draw (139) to (112);
		\draw (142) to (101);
		\draw (112) to (143);
		\draw (146.center) to (147);
		\draw (151.center) to (147);
		\draw (150.center) to (143);
		\draw (152) to (147);
		\draw [style=brace] (155.center) to (154.center);
		\draw [style=->] (162.center) to (163.center);
		\draw [style=->] (163.center) to (162.center);
		\draw (165) to (170.center);
		\draw (171.center) to (166);
		\draw (164) to (165);
		\draw (175) to (164);
		\draw (176) to (166);
	\end{pgfonlayer}
\end{tikzpicture}}
    \caption{HyperK\"ahler reduction for the hook case. The hyperK\"ahler quotients on the left are obtained by performing gaugings on the 3d mirrors on the right. }
    \label{fig:hook}
\end{figure}
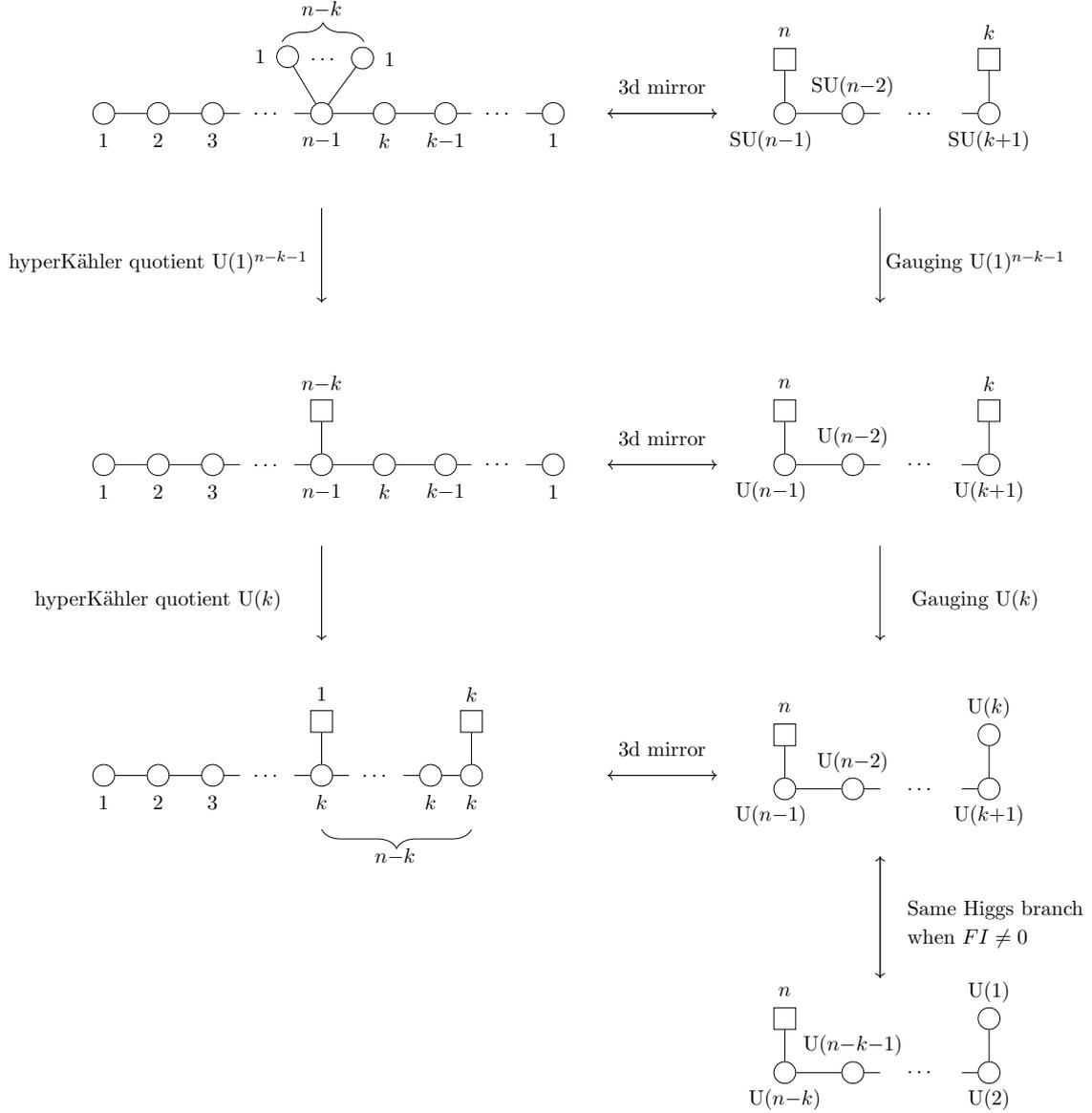

The action of taking the hyperK\"ahler quotient in \eqref{hookquiver} then translates to gauging flavour groups in its mirror, as shown in Figure \ref{fig:hook}. Following the gauging procedure of the mirror quiver, we arrived at the second quiver from the bottom right. This quiver clearly suffers from issues of incomplete Higgsing for all gauge nodes (except the first one from the left). However, the Higgs branch of a quiver suffering from incomplete Higgsing is equivalent to the Higgs branch of a different quiver with no Higgsing issues. To arrive at the mirror quiver on the bottom right, we use the prescription that  a $\mathrm{U}(k)$ gauge group with $N_f < 2k$ flavor has the same Higgs branch as a $\mathrm{U}(N_f-N_c)$ gauge group with $N_f$ flavors \cite{Yaakov:2013fza}. However, such a relation only holds when the FI parameters are non-zero. When the FI parameters are zero, the new quiver with the same Higgs branch takes a different form \cite{Assel:2017jgo}. This observation leads us to conclude that only when $FI \neq 0$ does the mirror quiver have the expected Higgs branch that is the closure of the $[n-k+1, 1^{k-1}] $  nilpotent orbit as predicted from the Levi group, see the bottom right part of Figure \ref{fig:hook}. 

Using the commutativity of the operations shown in Figure \ref{fig:hook}, one concludes that the HyperK\"ahler reduction of the Coulomb branch of (\ref{hookquiver}) by (\ref{leviHook}) is the expected nilpotent orbit (see the bottom right corner of Figure \ref{fig:implosionSummary2}), thus confirming the general claim (\ref{implodedQuiver}) for every hook partition.

\section{Orthosymplectic partial implosion}
\label{sec:orthosymplectic}

In this section, we look at some other classical groups. It is useful to recall that parabolic subgroups $P$, up to  conjugacy, are classified by subsets of the Dynkin diagram. The subdiagram is the Dynkin diagram of the commutator of the Levi subgroup $L_P$.\footnote{Note that the Levi subgroups of $\mathrm{SO}_{2n}$ are in one to one correspondence with the different phases of n Dp branes next to an Op${}^-$ plane.}
The partial flag varieties for $\mathrm{SO}_{2n}$ are of the form \footnote{Recall that $\mathrm{SO}_{2n}$ denotes a non-compact complex group, while $\mathrm{SO}(2n)$ denotes a compact real group.}
\begin{equation}
\frac{\mathrm{SO}_{2n}}{P}=
\frac{\mathrm{SO}(2n)}{\mathrm{U}(p_1) \times \ldots\times \mathrm{U}(p_r) \times \mathrm{SO}(2l)},
\end{equation}
where $P$ is a parabolic, defined by the decomposition 
\begin{equation}
\label{ParabolicsSO}
P \quad \leftrightarrow \quad l + \sum_{i=1}^{r} p_i = n,
\end{equation}
with $0 \leq l \leq n$. This reflects the fact that
subdiagrams of the $D_n$ Dynkin diagram have components of $A$ type, as well as possibly one $D$ type component.

It follows that $P$ has complex dimension 
$\frac{1}{2}(n(2n-1) + l(2l-1) + \sum_{i=1}^{r}p_i^2)$
and the quaternionic dimension of the partial implosion is
\begin{equation}
\label{dimPartialImplosionSO}
\mathrm{dim}_{\mathbb{H}} (T^\ast \mathrm{SO}_{2n})_{P,\mathrm{impl}} = \frac{1}{2} \left( n(2n-1) + l(2l-1) + \sum_{i=1}^{r} p_i^2 \right) \, .
\end{equation}
If $l=0$ and $p_i =1$ for all $i$ then we get the classical implosion with complex dimension $2n^2$.

We propose, in analogy with (\ref{implodedQuiver}), the following Coulomb branch description for the orthogonal universal partial implosion : 
\begin{equation}
\label{partialImplosionSO}
(T^\ast \mathrm{SO}_{2n})_{P,\mathrm{impl}} = \mathcal{C} \left( \raisebox{-.5\height}{\scalebox{.9}{
\begin{tikzpicture}
	\begin{pgfonlayer}{nodelayer}
		\node [style=redgauge] (0) at (-5, 0) {};
		\node [style=redgauge] (1) at (-3, 0) {};
		\node [style=none] (3) at (-2.5, 0) {};
		\node [style=none] (4) at (-1.25, 0) {};
		\node [style=redgauge] (5) at (-0.75, 0) {};
		\node [style=bluegauge] (6) at (-4, 0) {};
		\node [style=bluegauge] (7) at (0.25, 0) {};
		\node [style=none] (8) at (-5, -0.5) {2};
		\node [style=none] (9) at (-4, -0.5) {2};
		\node [style=none] (10) at (-3, -0.5) {4};
		\node [style=none] (11) at (-0.75, -0.5) {$2n{-}2$};
		\node [style=none] (12) at (0.25, -0.5) {$2n{-}2$};
		\node [style=gauge3] (13) at (1, 1) {};
		\node [style=none] (14) at (0.25, 1) {$p_1$};
		\node [style=none] (15) at (1.35, 1.5) {};
		\node [style=none] (16) at (2.025, 2.4) {};
		\node [style=gauge3] (17) at (2.375, 2.85) {};
		\node [style=gauge3] (18) at (3.05, 3.7) {};
		\node [style=none] (20) at (2, 2.875) {2};
		\node [style=none] (21) at (2.55, 3.725) {1};
		\node [style=gauge3] (22) at (1.5, -0.025) {};
		\node [style=none] (24) at (2.05, -0.025) {};
		\node [style=none] (25) at (3, -0.05) {};
		\node [style=gauge3] (26) at (3.5, -0.05) {};
		\node [style=gauge3] (27) at (4.6, -0.05) {};
		\node [style=none] (29) at (3.5, 0.55) {2};
		\node [style=none] (30) at (4.6, 0.525) {1};
		\node [style=bd] (31) at (3.825, 2.95) {};
		\node [style=bd] (32) at (4.475, 2.025) {};
		\node [style=bd] (33) at (4.75, 1) {};
		\node [style=none] (34) at (1.5, 0.45) {$p_r$};
		\node [style=redgauge] (35) at (1.25, -0.75) {};
		\node [style=none] (36) at (1.75, -0.75) {$2l$};
		\node [style=none] (37) at (1.775, -1.15) {};
		\node [style=none] (38) at (2.775, -1.95) {};
		\node [style=bluegauge] (39) at (3.25, -2.3) {};
		\node [style=redgauge] (40) at (3.9, -2.75) {};
		\node [style=bd] (41) at (1.475, 1.675) {};
		\node [style=bd] (42) at (1.675, 1.925) {};
		\node [style=bd] (43) at (1.875, 2.225) {};
		\node [style=bd] (44) at (2.275, -0.025) {};
		\node [style=bd] (45) at (2.525, -0.025) {};
		\node [style=bd] (46) at (2.775, -0.025) {};
		\node [style=bd] (47) at (1.975, -1.325) {};
		\node [style=bd] (49) at (2.25, -1.55) {};
		\node [style=bd] (50) at (2.55, -1.8) {};
		\node [style=none] (51) at (3.725, -2.275) {2};
		\node [style=none] (52) at (4.325, -2.775) {2};
		\node [style=bd] (53) at (-2.15, 0) {};
		\node [style=bd] (54) at (-1.9, 0) {};
		\node [style=bd] (55) at (-1.65, 0) {};
	\end{pgfonlayer}
	\begin{pgfonlayer}{edgelayer}
		\draw (0) to (6);
		\draw (6) to (1);
		\draw (1) to (3.center);
		\draw (4.center) to (5);
		\draw (5) to (7);
		\draw (7) to (13);
		\draw (17) to (18);
		\draw (17) to (16.center);
		\draw (13) to (15.center);
		\draw (26) to (27);
		\draw (26) to (25.center);
		\draw (22) to (24.center);
		\draw (7) to (22);
		\draw (7) to (35);
		\draw (35) to (37.center);
		\draw (38.center) to (39);
		\draw (39) to (40);
	\end{pgfonlayer}
\end{tikzpicture}}} \right)
\end{equation}
That is, we take the quiver for the $\mathrm{SO}_{2n}$ nilpotent cone,
and explode the flavour node into unitary legs
of length $p_i$ $(i=1, \ldots, r)$ 
and (if $l >0$) an orthosymplectic leg starting at
$\mathrm{SO}(2l)$. As in the unitary case, this keeps the
remaining nodes of the nilpotent quiver balanced, generating an $\mathrm{SO}(2n)$ symmetry.
The balanced nodes in the legs generate
$\mathrm{SU}(p_1) \times \mathrm{SU}(p_r) \times \mathrm{SO}(2l)$ symmetry, and the unbalanced nodes adjacent to the $\mathrm{USp}(2n-2)$
node give the remaining abelian symmetries.

Note that the balancing condition at the $\mathrm{SO}(2l)$ node of the orthosymplectic leg is
\begin{equation}
4l = (2l-2) + (2n-2) + 2 \, ,
\end{equation}
that is, $l = n-1$. 
This is the parabolic associated to the 
symmetric space coadjoint
orbit $\mathrm{SO}(2n)/\mathrm{U}(1) \times \mathrm{SO}(2n-2) =\mathrm{SO}(2n)/\mathrm{SO}(2) \times
\mathrm{SO}(2n-2)$, the hyperquadric in $\mathbb C \mathbb P^{2n-1}$. 
If this condition holds then there is a symmetry enhancement,
whereas if $l < n-1$ then the node has positive balance.

\medskip

As a first confirmation of the proposal, the rank of the gauge group defining the quiver in (\ref{partialImplosionSO}) is  
\begin{eqnarray}
\textrm{rank} &=& 2 \sum_{i=1}^{n-1} i + \sum_{i=1}^{r} 
\frac{1}{2} p_i (p_i+1) + l+ 2 \sum_{i=1}^{l-1} i \nonumber  \\
           &=& \frac{1}{2} \left( n(2n-1) + l(2l-1) + \sum_{i=1}^{r} p_i^2 \right)  \, , \nonumber 
\end{eqnarray}
which equals the quaternionic dimension (\ref{dimPartialImplosionSO}), as desired.

We now turn to the $\mathrm{SO}_8$ example to perform further checks.

\begin{figure}
    \centering
\scalebox{.9}{\begin{tikzpicture}
\node at (0,21.5) {Partitions};
\node at (4,21.5) { \begin{tabular}{c}
    Commutator \\
    of Levi
\end{tabular} };
\node at (7,21.5) {Diagrams};
\node (12) at (0,20) {$[7,1]$};
\node (11) at (0,18) {$[5,3]$};
\node (10a) at (-3,16) {$[4^2]_{\mathrm{I}}$};
\node (10b) at (-1,16) {$[4^2]_{\mathrm{II}}$};
\node (10c) at (2,14) {$[5,1^3]$};
\node (9) at (0,12) {$[3^2,1^2]$};
\node (8) at (0,10) {\textcolor{red}{$[3,2^2,1]$}};
\node (6a) at (2,8) {$[3,1^5]$};
\node (6b) at (-3,6) {$[2^4]_{\mathrm{I}}$};
\node (6c) at (-1,6) {$[2^4]_{\mathrm{II}}$};
\node (5) at (0,4) {\textcolor{red}{$[2^2,1^4]$}};
\node (0) at (0,2) {$[1^8]$};
\draw (12)--(11)--(10a)--(9)--(8)--(6a)--(5)--(0);
\draw (11)--(10b)--(9);
\draw (11)--(10c)--(9);
\draw (8)--(6a)--(5);
\draw (8)--(6b)--(5);
\draw (8)--(6c)--(5);
\node at (4,20) {$0$};
\node at (4,18) {$A_1$};
\node at (4,16) {$A_1+A_1$};
\node at (4,14) {$D_2$};
\node at (4,12.25) {$A_1+D_2$};
\node at (4,11.75) {$A_2$};
\node at (4,08) {$D_3$};
\node at (4,06) {$A_3$};
\node at (4,02) {$D_4$};
\node at (6,20) {\dfourquiver{white}{white}{white}{white}};
\node at (6,18) {\dfourquiver{white}{red}{white}{white}};
\node at (8,18) {\dfourquiver{white}{white}{red}{white}};
\node at (6,16) {\dfourquiver{red}{white}{red}{white}};
\node at (6,14) {\dfourquiver{white}{white}{red}{red}};
\node at (6,12) {\dfourquiver{red}{white}{red}{red}};
\node at (8,12) {\dfourquiver{white}{red}{red}{white}};
\node at (6,08) {\dfourquiver{white}{red}{red}{red}};
\node at (6,06) {\dfourquiver{red}{red}{red}{white}};
\node at (6,02) {\dfourquiver{red}{red}{red}{red}};
\end{tikzpicture}}
    \caption{Hasse diagram of $D_4$ nilpotent orbits, labeled by even integer partitions of 8. Partitions in black correspond to Richardson orbits, and partitions in red to non Richardson orbits. }
    \label{fig:nilpotentOrbitsD41}
\end{figure}
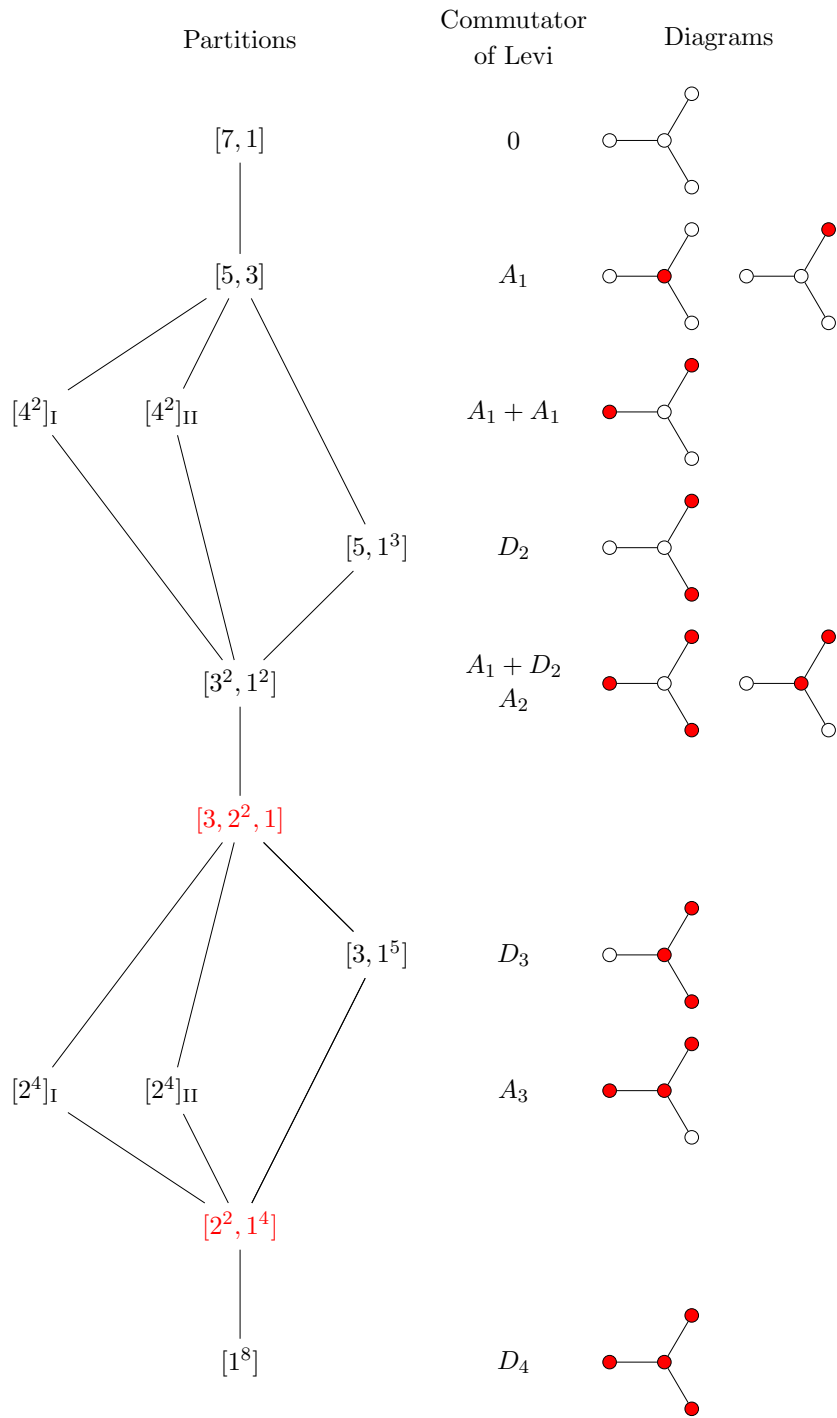

\subsection{The $\mathrm{SO}_8$ example}

There are 16 parabolic subgroups corresponding to subsets of the Dynkin diagram $D_4$, but some of these are equivalent under triality. 
The diagram in Figure \ref{fig:nilpotentOrbitsD41} displays the 12 nilpotent orbits, ten of which are Richardson orbits associated to parabolics. 
(The 'Levi' column gives the commutator of the Levi subgroup of the parabolic).

Some of these examples are equivalent under triality, for example
the $[4^2]_I$ and $[4^2]_{II}$ orbits and the $[5,1^3]$
orbit with Levi $D_2 \cong A_1 + A_1$. A similar
statement applies in the other case of a very even partition, the two $[2^4]$ orbits and the $[3,1^5]$ one.
Note that the $[3^2,1^2]$ orbit
comes from two distinct parabolics with
non-isomorphic Levi subgroups. The subregular orbit $[5,3]$ arises from parabolics with
isomorphic Levi but which are not related by triality (corresponding to the central node
or a peripheral node of the Dynkin diagram).

The other 2 orbits, given by partitions $[3,2^2,1]$ and the minimal orbit $[2^2,1^4]$, are not Richardson.

\begin{figure}
    \centering
\scalebox{.9}{\begin{tikzpicture}
\node at (0,21.5) {Partitions};
\node at (7,21.5) {Bouquets};
\node (12) at (0,20) {$[7,1]$};
\node (11) at (0,18) {$[5,3]$};
\node (10a) at (-3,16) {$[4^2]_{\mathrm{I}}$};
\node (10b) at (-1,16) {$[4^2]_{\mathrm{II}}$};
\node (10c) at (2,14) {$[5,1^3]$};
\node (9) at (0,12) {$[3^2,1^2]$};
\node (8) at (0,10) {\textcolor{red}{$[3,2^2,1]$}};
\node (6a) at (2,8) {$[3,1^5]$};
\node (6b) at (-3,6) {$[2^4]_{\mathrm{I}}$};
\node (6c) at (-1,6) {$[2^4]_{\mathrm{II}}$};
\node (5) at (0,4) {\textcolor{red}{$[2^2,1^4]$}};
\node (0) at (0,2) {$[1^8]$};
\draw (12)--(11)--(10a)--(9)--(8)--(6a)--(5)--(0);
\draw (11)--(10b)--(9);
\draw (11)--(10c)--(9);
\draw (8)--(6a)--(5);
\draw (8)--(6b)--(5);
\draw (8)--(6c)--(5);
\node at (6,20) {\scalebox{0.7}{\begin{tikzpicture}
	\begin{pgfonlayer}{nodelayer}
		\node [style=bluegauge] (0) at (0, 0) {};
		\node [style=gauge3] (1) at (1, 1.25) {};
		\node [style=gauge3] (2) at (1, 0.5) {};
		\node [style=gauge3] (3) at (1, -0.25) {};
		\node [style=redgauge] (4) at (0.75, -1) {};
		\node [style=none] (5) at (0, -0.5) {6};
		\node [style=none] (6) at (1.5, 1.25) {1};
		\node [style=none] (7) at (1.5, 0.5) {1};
		\node [style=none] (8) at (1.5, -0.25) {1};
		\node [style=none] (9) at (1, -1.5) {2};
		\node [style=none] (10) at (-0.5, 0) {};
		\node [style=none] (11) at (-1, 0) {$\dots$};
	\end{pgfonlayer}
	\begin{pgfonlayer}{edgelayer}
		\draw (0) to (1);
		\draw (2) to (0);
		\draw (0) to (3);
		\draw (0) to (4);
		\draw (0) to (10.center);
	\end{pgfonlayer}
\end{tikzpicture}
}
};
        \node at (6,18) {\scalebox{0.7}{\begin{tikzpicture}
	\begin{pgfonlayer}{nodelayer}
		\node [style=bluegauge] (0) at (0, 0) {};
		\node [style=gauge3] (1) at (1, 0.75) {};
		\node [style=gauge3] (2) at (1, 0) {};
		\node [style=gauge3] (3) at (1, -0.75) {};
		\node [style=none] (5) at (0, -0.5) {6};
		\node [style=none] (6) at (1, 1.25) {2};
		\node [style=none] (7) at (1.5, 0) {1};
		\node [style=none] (8) at (1.5, -0.75) {1};
		\node [style=none] (10) at (-0.5, 0) {};
		\node [style=none] (11) at (-1, 0) {$\dots$};
		\node [style=gauge3] (12) at (2, 0.75) {};
		\node [style=none] (13) at (2, 1.25) {1};
	\end{pgfonlayer}
	\begin{pgfonlayer}{edgelayer}
		\draw (0) to (1);
		\draw (2) to (0);
		\draw (0) to (3);
		\draw (0) to (10.center);
		\draw (1) to (12);
	\end{pgfonlayer}
\end{tikzpicture}
}};
\node at (6,16){\scalebox{0.7}{\begin{tikzpicture}
	\begin{pgfonlayer}{nodelayer}
		\node [style=bluegauge] (0) at (0, 0) {};
		\node [style=gauge3] (2) at (1, 1) {};
		\node [style=gauge3] (3) at (1, 0.25) {};
		\node [style=none] (5) at (0, -0.5) {6};
		\node [style=none] (7) at (1.5, 1) {1};
		\node [style=none] (8) at (1.5, 0.25) {1};
		\node [style=none] (10) at (-0.5, 0) {};
		\node [style=none] (11) at (-1, 0) {$\dots$};
		\node [style=redgauge] (12) at (1, -0.5) {};
		\node [style=bluegauge] (13) at (2, -0.5) {};
		\node [style=redgauge] (14) at (3, -0.5) {};
		\node [style=none] (15) at (1, -1) {4};
		\node [style=none] (16) at (2, -1) {2};
		\node [style=none] (17) at (3, -1) {2};
	\end{pgfonlayer}
	\begin{pgfonlayer}{edgelayer}
		\draw (2) to (0);
		\draw (0) to (3);
		\draw (0) to (10.center);
		\draw (0) to (12);
		\draw (12) to (13);
		\draw (13) to (14);
	\end{pgfonlayer}
\end{tikzpicture}
}};
\node at (6,14) {\scalebox{0.7}{\begin{tikzpicture}
	\begin{pgfonlayer}{nodelayer}
		\node [style=bluegauge] (0) at (0, 0) {};
		\node [style=gauge3] (2) at (1, 0.5) {};
		\node [style=gauge3] (3) at (1, -0.5) {};
		\node [style=none] (5) at (0, -0.5) {6};
		\node [style=none] (7) at (1, 1) {2};
		\node [style=none] (8) at (1, -1) {2};
		\node [style=none] (10) at (-0.5, 0) {};
		\node [style=none] (11) at (-1, 0) {$\dots$};
		\node [style=gauge3] (12) at (2, 0.5) {};
		\node [style=gauge3] (13) at (2, -0.5) {};
		\node [style=none] (14) at (2, 1) {1};
		\node [style=none] (15) at (2, -1) {1};
	\end{pgfonlayer}
	\begin{pgfonlayer}{edgelayer}
		\draw (2) to (0);
		\draw (0) to (3);
		\draw (0) to (10.center);
		\draw (2) to (12);
		\draw (13) to (3);
	\end{pgfonlayer}
\end{tikzpicture}
}};
\node at (6,12){\scalebox{0.7}{\begin{tikzpicture}
	\begin{pgfonlayer}{nodelayer}
		\node [style=bluegauge] (0) at (0, 0) {};
		\node [style=none] (5) at (0, -0.5) {6};
		\node [style=none] (10) at (-0.5, 0) {};
		\node [style=none] (11) at (-1, 0) {$\dots$};
		\node [style=redgauge] (12) at (1, -0.75) {};
		\node [style=gauge3] (13) at (1, 0.75) {};
		\node [style=gauge3] (14) at (2, 0.75) {};
		\node [style=gauge3] (15) at (3, 0.75) {};
		\node [style=none] (16) at (1, 1.25) {3};
		\node [style=none] (17) at (2, 1.25) {2};
		\node [style=none] (18) at (3, 1.25) {1};
		\node [style=none] (19) at (1.5, -0.75) {2};
	\end{pgfonlayer}
	\begin{pgfonlayer}{edgelayer}
		\draw (0) to (10.center);
		\draw (0) to (13);
		\draw (13) to (14);
		\draw (14) to (15);
		\draw (0) to (12);
	\end{pgfonlayer}
\end{tikzpicture}
}};
\node at (10,12){\scalebox{0.7}{\begin{tikzpicture}
	\begin{pgfonlayer}{nodelayer}
		\node [style=bluegauge] (0) at (0, 0) {};
		\node [style=gauge3] (2) at (1, 0.75) {};
		\node [style=none] (5) at (0, -0.5) {6};
		\node [style=none] (7) at (1, 1.25) {2};
		\node [style=none] (10) at (-0.5, 0) {};
		\node [style=none] (11) at (-1, 0) {$\dots$};
		\node [style=redgauge] (12) at (1, -0.5) {};
		\node [style=bluegauge] (13) at (2, -0.5) {};
		\node [style=redgauge] (14) at (3, -0.5) {};
		\node [style=none] (15) at (1, -1) {4};
		\node [style=none] (16) at (2, -1) {2};
		\node [style=none] (17) at (3, -1) {2};
		\node [style=gauge3] (18) at (2, 0.75) {};
		\node [style=none] (19) at (2, 1.25) {1};
	\end{pgfonlayer}
	\begin{pgfonlayer}{edgelayer}
		\draw (2) to (0);
		\draw (0) to (10.center);
		\draw (0) to (12);
		\draw (12) to (13);
		\draw (13) to (14);
		\draw (2) to (18);
	\end{pgfonlayer}
\end{tikzpicture}
}};
\node at (6,08) {\scalebox{0.7}{\begin{tikzpicture}
	\begin{pgfonlayer}{nodelayer}
		\node [style=bluegauge] (0) at (0, 0) {};
		\node [style=none] (5) at (0, -0.5) {6};
		\node [style=none] (10) at (-0.5, 0) {};
		\node [style=none] (11) at (-1, 0) {$\dots$};
		\node [style=gauge3] (13) at (1, 0) {};
		\node [style=gauge3] (14) at (2, 0) {};
		\node [style=gauge3] (15) at (3, 0) {};
		\node [style=none] (16) at (1, -0.5) {4};
		\node [style=none] (17) at (2, -0.5) {3};
		\node [style=none] (18) at (3, -0.5) {2};
		\node [style=gauge3] (19) at (4, 0) {};
		\node [style=none] (20) at (4, -0.5) {1};
	\end{pgfonlayer}
	\begin{pgfonlayer}{edgelayer}
		\draw (0) to (10.center);
		\draw (0) to (13);
		\draw (13) to (14);
		\draw (14) to (15);
		\draw (15) to (19);
	\end{pgfonlayer}
\end{tikzpicture}
}};
\node at (6,06) {\scalebox{0.7}{\begin{tikzpicture}
	\begin{pgfonlayer}{nodelayer}
		\node [style=bluegauge] (0) at (0, 0) {};
		\node [style=none] (5) at (0, -0.5) {6};
		\node [style=none] (10) at (-0.5, 0) {};
		\node [style=none] (11) at (-1, 0) {$\dots$};
		\node [style=gauge3] (13) at (1, 0.75) {};
		\node [style=none] (14) at (1, 1.25) {1};
		\node [style=redgauge] (15) at (1, -0.75) {};
		\node [style=redgauge] (16) at (3, -0.75) {};
		\node [style=redgauge] (17) at (5, -0.75) {};
		\node [style=bluegauge] (18) at (4, -0.75) {};
		\node [style=bluegauge] (19) at (2, -0.75) {};
		\node [style=none] (20) at (1, -1.25) {6};
		\node [style=none] (21) at (2, -1.25) {4};
		\node [style=none] (22) at (3, -1.25) {4};
		\node [style=none] (23) at (4, -1.25) {2};
		\node [style=none] (24) at (5, -1.25) {2};
	\end{pgfonlayer}
	\begin{pgfonlayer}{edgelayer}
		\draw (0) to (10.center);
		\draw (0) to (13);
		\draw (0) to (15);
		\draw (15) to (17);
	\end{pgfonlayer}
\end{tikzpicture}
}};
\node at (6,02) {\scalebox{0.7}{\begin{tikzpicture}
	\begin{pgfonlayer}{nodelayer}
		\node [style=bluegauge] (0) at (0, 0) {};
		\node [style=none] (5) at (0, -0.5) {6};
		\node [style=none] (10) at (-0.5, 0) {};
		\node [style=none] (11) at (-1, 0) {$\dots$};
		\node [style=redgauge] (15) at (1, 0) {};
		\node [style=redgauge] (16) at (3, 0) {};
		\node [style=redgauge] (17) at (5, 0) {};
		\node [style=bluegauge] (18) at (4, 0) {};
		\node [style=bluegauge] (19) at (2, 0) {};
		\node [style=none] (20) at (3, -0.5) {6};
		\node [style=none] (21) at (4, -0.5) {4};
		\node [style=none] (22) at (5, -0.5) {4};
		\node [style=none] (23) at (6, -0.5) {2};
		\node [style=none] (24) at (7, -0.5) {2};
		\node [style=bluegauge] (25) at (6, 0) {};
		\node [style=redgauge] (26) at (7, 0) {};
		\node [style=none] (27) at (2, -0.5) {6};
		\node [style=none] (28) at (1, -0.5) {8};
	\end{pgfonlayer}
	\begin{pgfonlayer}{edgelayer}
		\draw (0) to (10.center);
		\draw (0) to (15);
		\draw (15) to (17);
		\draw (17) to (26);
	\end{pgfonlayer}
\end{tikzpicture}
}};
\end{tikzpicture}}
    \caption{$D_4$ nilpotent orbits and the different bouquets of the $\mathrm{SO}(8)$ flavor node according to the partitions. }
    \label{fig:nilpotentOrbitsD42}
\end{figure}
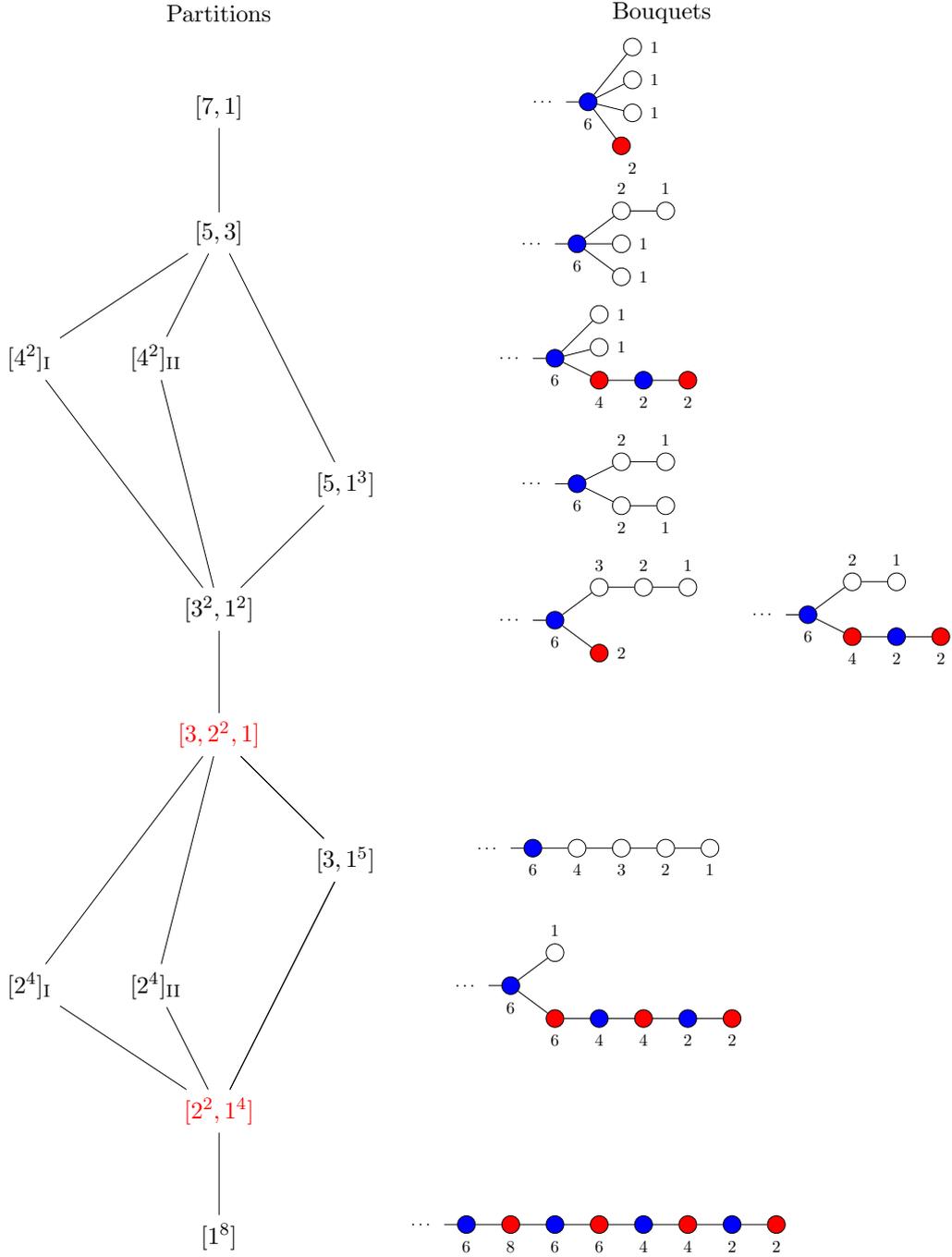

\paragraph{Subregular}
For the subregular case of $D_4$, the partition is $[5,3]$. This can be obtained by taking the following hyperK\"ahler quotient:
\begin{equation}
   \raisebox{-.5\height}{\scalebox{.8}{\begin{tikzpicture}
	\begin{pgfonlayer}{nodelayer}
		\node [style=redgauge] (0) at (-4.75, 0) {};
		\node [style=redgauge] (1) at (-2.75, 0) {};
		\node [style=redgauge] (5) at (-0.75, 0) {};
		\node [style=bluegauge] (6) at (-3.75, 0) {};
		\node [style=bluegauge] (7) at (0.25, 0) {};
		\node [style=none] (8) at (-4.75, -0.5) {2};
		\node [style=none] (9) at (-3.75, -0.5) {2};
		\node [style=none] (10) at (-2.75, -0.5) {4};
		\node [style=gauge3] (13) at (1, 1) {};
		\node [style=none] (14) at (0.5, 1) {2};
		\node [style=gauge3] (22) at (1.5, -0.025) {};
		\node [style=none] (34) at (1.5, 0.45) {1};
		\node [style=redgauge] (35) at (1, -1) {};
		\node [style=none] (56) at (1.5, -1) {2};
		\node [style=none] (57) at (-0.75, -0.5) {6};
		\node [style=none] (58) at (0.25, -0.5) {6};
		\node [style=bluegauge] (59) at (-1.75, 0) {};
		\node [style=none] (60) at (-1.75, -0.5) {4};
		\node [style=gauge3] (61) at (1.65, 1.875) {};
		\node [style=none] (62) at (1.175, 1.9) {1};
		\node [style=none] (63) at (3, 0) {};
		\node [style=none] (64) at (5, 0) {};
		\node [style=none] (65) at (4, 0.25) {hyperK\"ahler quotient};
		\node [style=redgauge] (66) at (6, 0) {};
		\node [style=redgauge] (67) at (8, 0) {};
		\node [style=redgauge] (68) at (10, 0) {};
		\node [style=bluegauge] (69) at (7, 0) {};
		\node [style=bluegauge] (70) at (11, 0) {};
		\node [style=none] (71) at (6, -0.5) {2};
		\node [style=none] (72) at (7, -0.5) {2};
		\node [style=none] (73) at (8, -0.5) {4};
		\node [style=none] (80) at (10, -0.5) {6};
		\node [style=none] (81) at (11, -0.5) {4};
		\node [style=bluegauge] (82) at (9, 0) {};
		\node [style=none] (83) at (9, -0.5) {4};
		\node [style=flavourBlue] (84) at (10, 1) {};
		\node [style=flavourRed] (85) at (11, 1) {};
		\node [style=none] (86) at (11, 1.5) {4};
		\node [style=none] (87) at (10, 1.5) {2};
	\end{pgfonlayer}
	\begin{pgfonlayer}{edgelayer}
		\draw (0) to (6);
		\draw (6) to (1);
		\draw (5) to (7);
		\draw (7) to (13);
		\draw (7) to (22);
		\draw (7) to (35);
		\draw (1) to (5);
		\draw (13) to (61);
		\draw [style=->] (63.center) to (64.center);
		\draw (66) to (69);
		\draw (69) to (67);
		\draw (68) to (70);
		\draw (67) to (68);
		\draw (84) to (68);
		\draw (85) to (70);
	\end{pgfonlayer}
\end{tikzpicture}}}
\end{equation}

Here, we look at an explicit computation of the Coulomb branch Hilbert series with $n=4$ with Jordan type $[5,3]$. The quiver is drawn with the following fugacities:
\begin{equation}
 \raisebox{-.5\height}{\begin{tikzpicture}
	\begin{pgfonlayer}{nodelayer}
		\node [style=redgauge] (0) at (-4.75, 0) {};
		\node [style=redgauge] (1) at (-2.75, 0) {};
		\node [style=redgauge] (5) at (-0.75, 0) {};
		\node [style=bluegauge] (6) at (-3.75, 0) {};
		\node [style=bluegauge] (7) at (0.25, 0) {};
		\node [style=none] (8) at (-4.75, -0.5) {2};
		\node [style=none] (9) at (-3.75, -0.5) {2};
		\node [style=none] (10) at (-2.75, -0.5) {4};
		\node [style=gauge3] (13) at (1, 1) {};
		\node [style=none] (14) at (0.5, 1) {2};
		\node [style=gauge3] (22) at (1.5, -0.025) {};
		\node [style=none] (34) at (1.5, 0.45) {1};
		\node [style=redgauge] (35) at (1, -1) {};
		\node [style=none] (56) at (1.5, -1) {2};
		\node [style=none] (57) at (-0.75, -0.5) {6};
		\node [style=none] (58) at (0.25, -0.5) {6};
		\node [style=bluegauge] (59) at (-1.75, 0) {};
		\node [style=none] (60) at (-1.75, -0.5) {4};
		\node [style=gauge3] (61) at (1.65, 1.875) {};
		\node [style=none] (62) at (1.175, 1.9) {1};
		\node [style=none] (66) at (2, -1) {$\color{red}{z_2}$};
		\node [style=none] (67) at (1.5, 1) {$\color{red}{q}$};
		\node [style=none] (68) at (2.175, 1.875) {$\color{red}{x}$};
		\node [style=none] (69) at (1.975, -0.025) {$\color{red}{z_1}$};
	\end{pgfonlayer}
	\begin{pgfonlayer}{edgelayer}
		\draw (0) to (6);
		\draw (6) to (1);
		\draw (5) to (7);
		\draw (7) to (13);
		\draw (7) to (22);
		\draw (7) to (35);
		\draw (1) to (5);
		\draw (13) to (61);
	\end{pgfonlayer}
\end{tikzpicture}}
\label{beforequotientSO}
\end{equation}
where $z_1,z_2,q$ are the fugacities of the three $\mathrm{U}(1)$ subgroups and $x$ is the fugacity of the $\mathrm{SU}(2)$ subgroup. The Coulomb branch Hilbert series is $\widetilde{\mathrm{HS}}_{[5,3]}(z_1,z_2,q,x;t) $. The computation of the Hilbert series requires the summation of dressed monopole operators with integer magnetic charges and half-plus integer magnetic charges (see \cite{Bourget:2020xdz}). The hyperK\"ahler quotient takes $\widetilde{\mathrm{HS}}_{[5,3]}(z_1,z_2,q,x;t) $ to $\mathrm{HS}_{[5,3]}(t) $:
\begin{equation}
\begin{split}
   \oint \frac{\, \mathrm{d}z_1\, \mathrm{d}z_2 \, \mathrm{d} q \, \mathrm{d} x}{(2\pi i)^4 z_1 z_2 q}   & \frac{1-x^2}{x} (1-t^2)^4(1-x^2t^2)(1-\frac{t^2}{x^2})\widetilde{\mathrm{HS}}_{[5,3]}(z_1,z_2,q,x;t) \\=  & 1+28t^2+405t^4\dots = \mathrm{HS}_{[5,3]}(t) 
      \end{split}
\end{equation}
where
\begin{equation}
\begin{split}
  \widetilde{\mathrm{HS}}_{[5,3]}(z_1,z_2,q,x;t) & =  1+\left(32+\frac{1}{x^2}+x^2\right )t^2\\ +& \left(528+\frac{1}{x^4}+ \frac{32}{x^2}+32x^2+x^4\right)t^4 +\dots
   \end{split}
\end{equation}
Note again the $t^2$ coefficient in $\mathrm{HS}_{[5,3]}$ gives the dimension of the global symmetry group $SO_8$.

\subsection{The $E_6$ family}
\label{sec:E6family}

For the unitary quivers, the hook quivers offer a nice family for us to test the hyperK\"ahler quotients using their mirror theories. However, for partial implosion of orthosymplectic quivers, it is more difficult to find a family of quivers with known mirror duals. 

In this section we look at a particular orthosymplectic example that has
interesting links with another space.
A family of moduli spaces, denoted $E_6^{(k)}$, defined by \cite{Bourget:2020gzi}
\begin{eqnarray}
    E_6^{(k)} &:=& \mathcal{H}^{5d,g_{YM} \rightarrow\infty} \left( \raisebox{-.5\height}{\begin{tikzpicture}
	\node (g1) [gauge,label=below:{$\mathrm{SU}(k{+}1)_{\pm \frac{3}{2}}$}] {};
	\node (g2) [flavour,above of=g1,label=above:{$2k{+}3$}] {};
	\draw (g1)--(g2);
	\end{tikzpicture}} \right) =  \mathcal{H}^{5d,g_{YM} \rightarrow\infty} \left( \raisebox{-.5\height}{\begin{tikzpicture}
	\node (g1) [gauge, fill=blue, label=below:{$\mathrm{Sp}(k)$}] {};
	\node (g2) [flavour, fill=red, above of=g1,label=above:{$D_{2k+3}$}] {};
	\draw (g1)--(g2);
	\end{tikzpicture}} \right) \nonumber \\
	&=& \mathcal{C}^{3d} \left( \raisebox{-.5\height}{\begin{tikzpicture}
	\tikzset{node distance = .8cm};
	\node (g1) [gauge,label=below:{\footnotesize{\rotatebox[origin=c]{0}{$1$}}}] {};
	\node (g2) [right of=g1] {$\cdots$};
	\node (g3) [gauge,right of=g2,label=below:{\footnotesize{\rotatebox[origin=c]{-45}{$2k+1$}}}] {};
	\node (g4) [gauge,right of=g3,label=below:{\footnotesize{\rotatebox[origin=c]{-45}{$k+1$}}}] {};
	\node (g5) [gauge,right of=g4,label=below:{\footnotesize{\rotatebox[origin=c]{0}{$1$}}}] {};
	\node (g7) [gauge,above of=g3,label=left:{\footnotesize{$k+1$}}] {};
	\node (g8) [gauge,right of=g7,label=right:{\footnotesize{$1$}}] {};
	\draw (g1)--(g2)--(g3)--(g4)--(g5) (g3)--(g7)--(g8);
	\end{tikzpicture}} \right) \\
	&=& \mathcal{C}^{3d} \left( \raisebox{-.5\height}{\begin{tikzpicture}
	\tikzset{node distance = .8cm};
	\node (g1) [gauge, fill=red, label=below:{\footnotesize{2}}] {};
	\node (g2) [gauge, right of=g1, fill=blue, label=below:{\footnotesize{2}}] {};
	\node (g3) [right of=g2] {$\cdots$};
	\node (g4) [gauge, right of=g3, fill=red, label=below:{\footnotesize{\rotatebox[origin=c]{-45}{$2k+2$}}}] {};
	\node (g5) [gauge, right of=g4, fill=blue, label=below:{\footnotesize{\rotatebox[origin=c]{-45}{$2k+2$}}}] {};
	\node (g6) [gauge, right of=g5, fill=red, label=below:{\footnotesize{\rotatebox[origin=c]{-45}{$2k+2$}}}] {};
	\node (g7) [right of=g6] {$\cdots$};
	\node (g8) [gauge, right of=g7, fill=blue, label=below:{\footnotesize{2}}] {};
	\node (g9) [gauge, right of=g8, fill=red, label=below:{\footnotesize{2}}] {};
	\node (g10) [gauge, above of=g5, label=above:{\footnotesize{1}}] {};
	\draw  (g1)--(g2)--(g3)--(g4)--(g5)--(g6)--(g7)--(g8)--(g9) (g5)--(g10);
	\end{tikzpicture}} \right) \nonumber
\end{eqnarray}
That is, we have the conjectured partial
implosion for $\mathrm{SO}(2k+4)$, with, in the notation of (\ref{ParabolicsSO}),
\begin{equation}
n = k+2, \; l = k+1, \; r=1, \; p_1 =1.
\end{equation}
As $l= n-1$ the $\mathrm{SO}(2l) = D_{k+1}$ node is now balanced and we get a global
symmetry enhancement from $\mathrm{SO}(2k+4) \times \mathrm{SO}(2k+2) \times \mathrm{U}(1)$
to $\mathrm{SO}(4k+6) \times \mathrm{U}(1)$. Consider the Levi subgroup
$\mathrm{SO}(2k+2) \times \mathrm{U}(1) \cong
\mathrm{SO}(2k+2) \times \mathrm{SO}(2)$ and take the hyperK\"ahler quotient:
\begin{equation}
\label{HKE6}
    X^{(k)} = E_6^{(k)} \tripleslash \left( \mathrm{SO}(2k+2) \times \mathrm{SO}(2) \right).
\end{equation}
We conjecture that $X^{(k)}$ is the closure of the next to minimal nilpotent orbit of $\mathrm{D}_{k+2}$ for $k \geq 1$. This conjecture leads to 
\begin{equation}
\label{ConjectureE6}
    \tilde{X}^{(k)} = T^{\ast} \left( \frac{\mathrm{SO}(2k+4)}{\mathrm{SO}(2k+2) \times \mathrm{SO}(2)} \right),
\end{equation}
where $\tilde{X}^{(k)}$ is the Springer resolution of $X^{(k)}$. One can check that the dimensions of the various spaces involved are in agreement with this conjecture. We would like to make more checks.

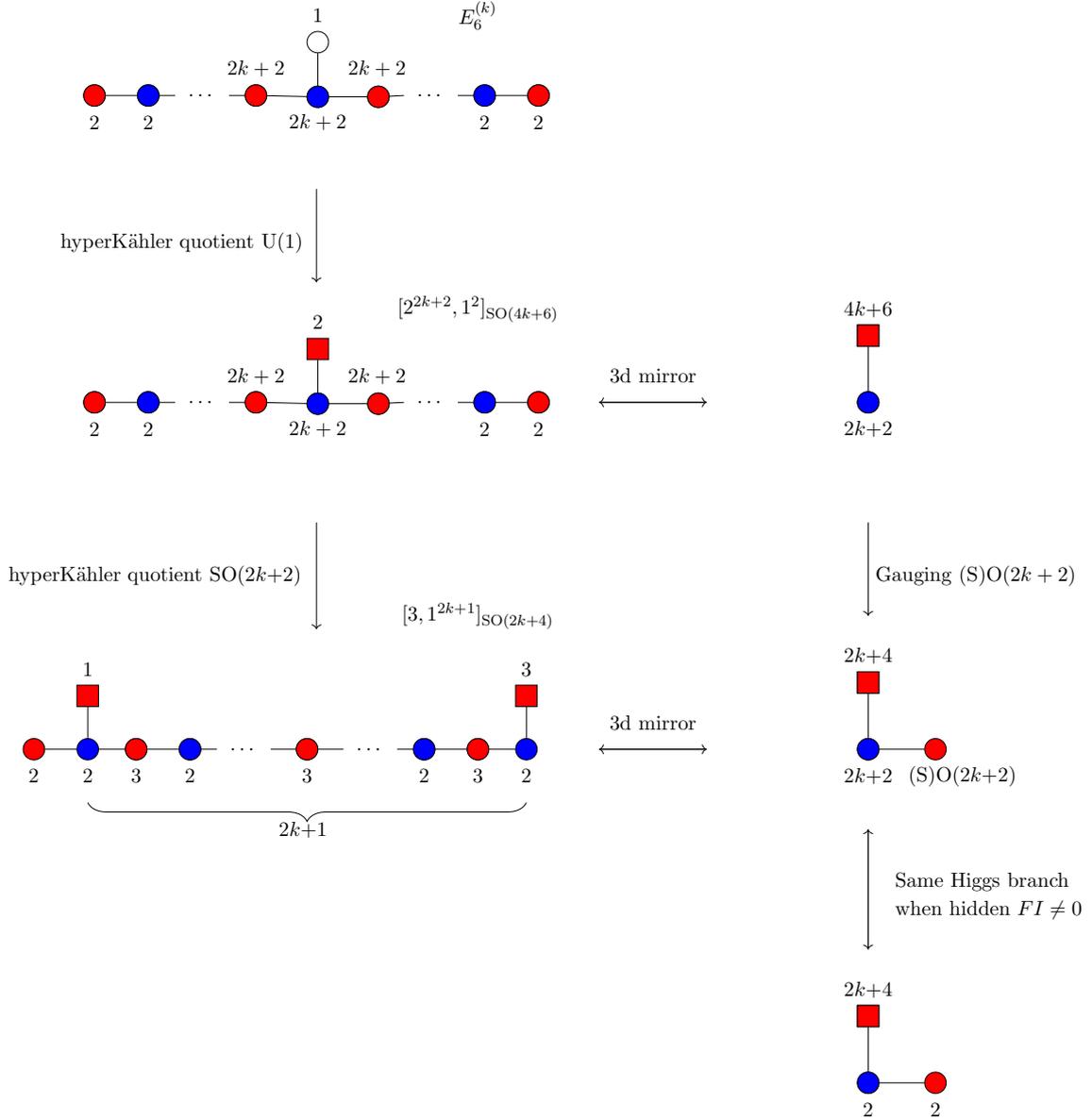
\begin{figure}
    \centering
\scalebox{0.75}{
\begin{tikzpicture}
	\begin{pgfonlayer}{nodelayer}
		\node [style=gauge3] (0) at (4.625, 0) {};
		\node [style=gauge3] (1) at (5.775, -0.025) {};
		\node [style=gauge3] (2) at (6.9, -0.025) {};
		\node [style=none] (3) at (4.125, 0) {};
		\node [style=none] (4) at (3.125, 0) {};
		\node [style=gauge3] (5) at (2.625, 0) {};
		\node [style=gauge3] (6) at (1.625, 0) {};
		\node [style=none] (7) at (7.875, 0) {\dots};
		\node [style=gauge3] (8) at (8.875, 0) {};
		\node [style=none] (9) at (7.375, 0) {};
		\node [style=none] (10) at (8.375, 0) {};
		\node [style=gauge3] (11) at (16, -5.75) {};
		\node [style=none] (12) at (16, -6.25) {$2k{+}2$};
		\node [style=none] (13) at (16, -4) {$4k{+}6$};
		\node [style=flavour2] (14) at (16, -4.5) {};
		\node [style=none] (15) at (11, -5.75) {};
		\node [style=none] (16) at (13, -5.75) {};
		\node [style=none] (17) at (5.75, -1.75) {};
		\node [style=none] (18) at (5.75, -3.5) {};
		\node [style=none] (19) at (3.25, -2.75) {hyperK\"ahler quotient $\mathrm{U}(1)$};
		\node [style=none] (20) at (11, -12.25) {};
		\node [style=none] (21) at (13, -12.25) {};
		\node [style=none] (22) at (5.75, -8) {};
		\node [style=none] (23) at (5.75, -10) {};
		\node [style=none] (24) at (16, -8) {};
		\node [style=none] (25) at (16, -9.75) {};
		\node [style=none] (26) at (2.75, -9) {hyperK\"ahler quotient $\mathrm{SO}(2k{+}2)$};
		\node [style=none] (27) at (18, -9) {Gauging $\mathrm{(S)O}(2k+2)$};
		\node [style=none] (28) at (12, -5.25) {3d mirror};
		\node [style=none] (29) at (12, -11.75) {3d mirror};
		\node [style=none] (30) at (3.625, 0) {\dots};
		\node [style=none] (31) at (16, -13.75) {};
		\node [style=none] (32) at (16, -16) {};
		\node [style=gauge3] (33) at (9.875, 0) {};
		\node [style=gauge3] (34) at (5.775, 1) {};
		\node [style=bluegauge] (35) at (2.625, 0) {};
		\node [style=bluegauge] (36) at (8.875, 0) {};
		\node [style=bluegauge] (37) at (5.775, -0.025) {};
		\node [style=redgauge] (38) at (9.875, 0) {};
		\node [style=redgauge] (39) at (6.9, -0.025) {};
		\node [style=redgauge] (40) at (4.625, 0) {};
		\node [style=redgauge] (41) at (1.625, 0) {};
		\node [style=none] (42) at (5.775, 1.5) {1};
		\node [style=none] (43) at (5.775, -0.5) {$2k+2$};
		\node [style=none] (44) at (6.875, 0.5) {$2k+2$};
		\node [style=none] (45) at (4.625, 0.5) {$2k+2$};
		\node [style=none] (46) at (1.625, -0.5) {2};
		\node [style=none] (47) at (2.625, -0.5) {2};
		\node [style=none] (48) at (8.875, -0.5) {2};
		\node [style=none] (49) at (9.875, -0.5) {2};
		\node [style=gauge3] (50) at (4.625, -5.75) {};
		\node [style=gauge3] (51) at (5.775, -5.775) {};
		\node [style=gauge3] (52) at (6.9, -5.775) {};
		\node [style=none] (53) at (4.125, -5.75) {};
		\node [style=none] (54) at (3.125, -5.75) {};
		\node [style=gauge3] (55) at (2.625, -5.75) {};
		\node [style=gauge3] (56) at (1.625, -5.75) {};
		\node [style=none] (57) at (7.875, -5.75) {\dots};
		\node [style=gauge3] (58) at (8.875, -5.75) {};
		\node [style=none] (59) at (7.375, -5.75) {};
		\node [style=none] (60) at (8.375, -5.75) {};
		\node [style=none] (61) at (3.625, -5.75) {\dots};
		\node [style=gauge3] (62) at (9.875, -5.75) {};
		\node [style=bluegauge] (63) at (2.625, -5.75) {};
		\node [style=bluegauge] (64) at (8.875, -5.75) {};
		\node [style=bluegauge] (65) at (5.775, -5.775) {};
		\node [style=redgauge] (66) at (9.875, -5.75) {};
		\node [style=redgauge] (67) at (6.9, -5.775) {};
		\node [style=redgauge] (68) at (4.625, -5.75) {};
		\node [style=redgauge] (69) at (1.625, -5.75) {};
		\node [style=none] (70) at (5.775, -4.25) {2};
		\node [style=none] (71) at (5.775, -6.25) {$2k+2$};
		\node [style=none] (72) at (6.875, -5.25) {$2k+2$};
		\node [style=none] (73) at (4.625, -5.25) {$2k+2$};
		\node [style=none] (74) at (1.625, -6.25) {2};
		\node [style=none] (75) at (2.625, -6.25) {2};
		\node [style=none] (76) at (8.875, -6.25) {2};
		\node [style=none] (77) at (9.875, -6.25) {2};
		\node [style=flavourRed] (78) at (5.775, -4.75) {};
		\node [style=none] (79) at (4.9, -12.25) {};
		\node [style=none] (80) at (3.9, -12.25) {};
		\node [style=gauge3] (81) at (3.4, -12.25) {};
		\node [style=gauge3] (82) at (2.4, -12.25) {};
		\node [style=none] (83) at (4.4, -12.25) {\dots};
		\node [style=bluegauge] (84) at (3.4, -12.25) {};
		\node [style=redgauge] (85) at (2.4, -12.25) {};
		\node [style=none] (86) at (2.4, -12.75) {3};
		\node [style=none] (87) at (3.4, -12.75) {2};
		\node [style=redgauge] (88) at (0.5, -12.25) {};
		\node [style=bluegauge] (89) at (1.5, -12.25) {};
		\node [style=none] (90) at (0.5, -12.75) {2};
		\node [style=none] (91) at (1.5, -12.75) {2};
		\node [style=flavourRed] (92) at (1.5, -11.25) {};
		\node [style=none] (93) at (1.5, -10.75) {1};
		\node [style=none] (94) at (6.25, -12.25) {};
		\node [style=none] (95) at (7.25, -12.25) {};
		\node [style=gauge3] (96) at (7.75, -12.25) {};
		\node [style=gauge3] (97) at (8.75, -12.25) {};
		\node [style=none] (98) at (6.75, -12.25) {\dots};
		\node [style=bluegauge] (99) at (7.75, -12.25) {};
		\node [style=redgauge] (100) at (8.75, -12.25) {};
		\node [style=none] (101) at (8.75, -12.75) {3};
		\node [style=none] (102) at (7.75, -12.75) {2};
		\node [style=bluegauge] (103) at (9.65, -12.25) {};
		\node [style=none] (104) at (9.65, -12.75) {2};
		\node [style=flavourRed] (105) at (9.65, -11.25) {};
		\node [style=none] (106) at (9.65, -10.75) {3};
		\node [style=gauge3] (107) at (5.575, -12.25) {};
		\node [style=redgauge] (108) at (5.575, -12.25) {};
		\node [style=none] (109) at (5.575, -12.75) {3};
		\node [style=none] (110) at (1.5, -13.25) {};
		\node [style=none] (111) at (9.65, -13.25) {};
		\node [style=none] (112) at (5.5, -13.75) {$2k{+}1$};
		\node [style=bluegauge] (113) at (16, -5.75) {};
		\node [style=flavourRed] (114) at (16, -4.5) {};
		\node [style=gauge3] (115) at (16, -12.25) {};
		\node [style=none] (116) at (16, -12.75) {$2k{+}2$};
		\node [style=none] (117) at (16, -10.5) {$2k{+}4$};
		\node [style=flavour2] (118) at (16, -11) {};
		\node [style=bluegauge] (119) at (16, -12.25) {};
		\node [style=flavourRed] (120) at (16, -11) {};
		\node [style=redgauge] (121) at (17.25, -12.25) {};
		\node [style=none] (122) at (17.75, -12.75) {$\mathrm{(S)O}(2k{+}2)$};
		\node [style=gauge3] (123) at (16, -18.5) {};
		\node [style=none] (124) at (16, -19) {2};
		\node [style=none] (125) at (16, -16.75) {$2k{+}4$};
		\node [style=flavour2] (126) at (16, -17.25) {};
		\node [style=bluegauge] (127) at (16, -18.5) {};
		\node [style=flavourRed] (128) at (16, -17.25) {};
		\node [style=redgauge] (129) at (17.25, -18.5) {};
		\node [style=none] (130) at (17.25, -19) {2};
		\node [style=none] (131) at (8.75, 1.5) {$E^{(k)}_6$};
		\node [style=none] (132) at (8.75, -4) {$[2^{2k{+}2},1^2]_{\mathrm{SO}(4k{+}6)}$};
		\node [style=none] (133) at (8.75, -9.75) {$[3,1^{2k{+}1}]_{\mathrm{SO}(2k{+}4)}$};
		\node [style=none] (134) at (19, -15) {\parbox{5cm}{Same Higgs branch\\ when hidden $FI \neq 0$}};
	\end{pgfonlayer}
	\begin{pgfonlayer}{edgelayer}
		\draw (0) to (1);
		\draw (1) to (2);
		\draw (5) to (4.center);
		\draw (6) to (5);
		\draw (3.center) to (0);
		\draw (2) to (9.center);
		\draw (10.center) to (8);
		\draw [style=->] (15.center) to (16.center);
		\draw [style=->] (16.center) to (15.center);
		\draw [style=->] (17.center) to (18.center);
		\draw [style=->] (20.center) to (21.center);
		\draw [style=->] (21.center) to (20.center);
		\draw [style=->] (22.center) to (23.center);
		\draw [style=->] (24.center) to (25.center);
		\draw [style=->] (31.center) to (32.center);
		\draw [style=->] (32.center) to (31.center);
		\draw (8) to (33);
		\draw (34) to (37);
		\draw (50) to (51);
		\draw (51) to (52);
		\draw (55) to (54.center);
		\draw (56) to (55);
		\draw (53.center) to (50);
		\draw (52) to (59.center);
		\draw (60.center) to (58);
		\draw (58) to (62);
		\draw (78) to (65);
		\draw (81) to (80.center);
		\draw (82) to (81);
		\draw (89) to (85);
		\draw (88) to (89);
		\draw (92) to (89);
		\draw (96) to (95.center);
		\draw (97) to (96);
		\draw (103) to (100);
		\draw (105) to (103);
		\draw (108) to (79.center);
		\draw (108) to (94.center);
		\draw [style=brace] (111.center) to (110.center);
		\draw (119) to (121);
		\draw (114) to (113);
		\draw (120) to (119);
		\draw (127) to (129);
		\draw (128) to (127);
	\end{pgfonlayer}
\end{tikzpicture}
}
    \caption{HyperK\"ahler reduction of $E_6^{(k)}$. The hyperK\"ahler quotients on the left are obtained by performing gaugings on the 3d mirrors on the right.}
    \label{fig:HKexplicit}
\end{figure}

Note that $X^{(k)}$ has an $\mathrm{SO}_{2k+4}$ action with complex moment map $\mu_{\mathbb C}$ valued in $\frak{so}_{2k+4}$. Equivariance of the moment map means that orbits are mapped onto coadjoint orbits in $\frak{so}_{2k+4}$. So the moment map is either a diffeomorphism onto the next-to-minimal orbit or maps orbits onto the minimal orbit.

We write out the hyperK\"ahler quotient (\ref{HKE6}) explicitly in Figure \ref{fig:HKexplicit}, following the reasoning pictured in Figure \ref{fig:hook}. The Coulomb branch of the quivers on the left (from top to bottom) are the $E^{k}_6$, $[2^{2k+2},1^2]$ orbit closure of $\mathrm{SO}_{4k+6}$ and $[3,1^{2k+1}]$ orbit closure of $\mathrm{SO}_{2k+4}$ respectively. The last equality uses the fact that the (hidden) FI parameters for the theories on the right column are taken to be non-zero, using the results recalled in Section \ref{sec:phys}. In conclusion, we have obtained a proof of the claim (\ref{ConjectureE6}), which is an argument in favor of our general conjecture (\ref{partialImplosionSO}). 

\section*{Conclusion}

We have provided a conjectured realization of universal partial implosions for $G = \mathrm{SL}_n$ and  $G = \mathrm{SO}_{2n}$ in terms of Coulomb branches, respectively in equations (\ref{implodedQuiver}) and (\ref{partialImplosionSO}).  In both cases, we have been able to check the conjectures by dimension computations and explicit calculations for certain parabolics. The phenomenon of non-complete Higgsing is responsible for the difficulty in proving the conjectures in the general case. We also leave for future work the determination of universal partial implosions for other simple groups.

\section*{Acknowledgements}
We are grateful to Frances Kirwan for useful discussions.
The work of A.~B.~, J.~F.~G.~, A.~H.~ and Z.~Z.~ is supported by STFC grants
ST/P000762/1 and ST/T000791/1. A.~B.~ is supported by the ERC Consolidator Grant 772408-Stringlandscape, and by the LabEx ENS-ICFP: ANR-10-LABX-0010/ANR-10-IDEX-0001-02 PSL*. 
A.~H.~ would like to thank the Isaac Newton Institute for Mathematical Sciences, Cambridge, for support and hospitality during the programme Cluster algebras and representation theory (CAR) where partial work on this paper was undertaken. This work was supported by EPSRC grant no EP/R014604/1.

\bibliographystyle{JHEP}
\bibliography{bibli.bib}

\providecommand{\href}[2]{#2}\begingroup\raggedright\begin{thebibliography}{10}

\bibitem{DHK}
A.~{Dancer}, A.~{Hanany} and F.~{Kirwan}, \emph{{ Symplectic duality and
  implosions}},  \href{https://arxiv.org/abs/2004.09620}{{\ttfamily
  2004.09620}}.

\bibitem{Bourget:2021zyc}
A.~Bourget, A.~Dancer, J.~F. Grimminger, A.~Hanany, F.~Kirwan and Z.~Zhong,
  \emph{{Orthosymplectic implosions}},
  \href{http://dx.doi.org/10.1007/JHEP08(2021)012}{\emph{JHEP} {\bfseries 08}
  (2021) 012}, [\href{https://arxiv.org/abs/2103.05458}{{\ttfamily
  2103.05458}}].

\bibitem{Hitchin:1986ea}
N.~J. Hitchin, A.~Karlhede, U.~Lindstrom and M.~Rocek, \emph{{Hyperkahler
  Metrics and Supersymmetry}},
  \href{http://dx.doi.org/10.1007/BF01214418}{\emph{Commun. Math. Phys.}
  {\bfseries 108} (1987) 535}.

\bibitem{Cremonesi:2013lqa}
S.~Cremonesi, A.~Hanany and A.~Zaffaroni, \emph{{Monopole operators and Hilbert
  series of Coulomb branches of $3d$ $\mathcal{N} = 4$ gauge theories}},
  \href{http://dx.doi.org/10.1007/JHEP01(2014)005}{\emph{JHEP} {\bfseries 01}
  (2014) 005}, [\href{https://arxiv.org/abs/1309.2657}{{\ttfamily 1309.2657}}].

\bibitem{Bullimore:2015lsa}
M.~Bullimore, T.~Dimofte and D.~Gaiotto, \emph{{The Coulomb Branch of 3d
  ${\mathcal{N}= 4}$ Theories}},
  \href{http://dx.doi.org/10.1007/s00220-017-2903-0}{\emph{Commun. Math. Phys.}
  {\bfseries 354} (2017) 671--751},
  [\href{https://arxiv.org/abs/1503.04817}{{\ttfamily 1503.04817}}].

\bibitem{Nakajima:2015txa}
H.~Nakajima, \emph{{Towards a mathematical definition of Coulomb branches of
  $3$-dimensional $\mathcal{N}=4$ gauge theories, I}},
  \href{http://dx.doi.org/10.4310/ATMP.2016.v20.n3.a4}{\emph{Adv. Theor. Math.
  Phys.} {\bfseries 20} (2016) 595--669},
  [\href{https://arxiv.org/abs/1503.03676}{{\ttfamily 1503.03676}}].

\bibitem{Braverman:2016wma}
A.~Braverman, M.~Finkelberg and H.~Nakajima, \emph{{Towards a mathematical
  definition of Coulomb branches of $3$-dimensional $\mathcal N=4$ gauge
  theories, II}}, {\emph{Adv. Theor.Math. Phys.} {\bfseries 22} (2018)
  1071--1147}, [\href{https://arxiv.org/abs/1601.03586}{{\ttfamily
  1601.03586}}].

\bibitem{Bourget:2021jwo}
A.~Bourget, J.~F. Grimminger, A.~Hanany, R.~Kalveks and Z.~Zhong, \emph{{Higgs
  Branches of U/SU Quivers via Brane Locking}},
  \href{https://arxiv.org/abs/2111.04745}{{\ttfamily 2111.04745}}.

\bibitem{Yaakov:2013fza}
I.~Yaakov, \emph{{Redeeming Bad Theories}},
  \href{http://dx.doi.org/10.1007/JHEP11(2013)189}{\emph{JHEP} {\bfseries 11}
  (2013) 189}, [\href{https://arxiv.org/abs/1303.2769}{{\ttfamily 1303.2769}}].

\bibitem{Assel:2017jgo}
B.~Assel and S.~Cremonesi, \emph{{The Infrared Physics of Bad Theories}},
  \href{http://dx.doi.org/10.21468/SciPostPhys.3.3.024}{\emph{SciPost Phys.}
  {\bfseries 3} (2017) 024},
  [\href{https://arxiv.org/abs/1707.03403}{{\ttfamily 1707.03403}}].

\bibitem{Bourget:2019aer}
A.~Bourget, S.~Cabrera, J.~F. Grimminger, A.~Hanany, M.~Sperling, A.~Zajac
  et~al., \emph{{The Higgs mechanism \textemdash{} Hasse diagrams for
  symplectic singularities}},
  \href{http://dx.doi.org/10.1007/JHEP01(2020)157}{\emph{JHEP} {\bfseries 01}
  (2020) 157}, [\href{https://arxiv.org/abs/1908.04245}{{\ttfamily
  1908.04245}}].

\bibitem{Seiberg:1994pq}
N.~Seiberg, \emph{{Electric - magnetic duality in supersymmetric nonAbelian
  gauge theories}},
  \href{http://dx.doi.org/10.1016/0550-3213(94)00023-8}{\emph{Nucl. Phys. B}
  {\bfseries 435} (1995) 129--146},
  [\href{https://arxiv.org/abs/hep-th/9411149}{{\ttfamily hep-th/9411149}}].

\bibitem{Aharony:1997gp}
O.~Aharony, \emph{{IR duality in d = 3 N=2 supersymmetric USp(2N(c)) and
  U(N(c)) gauge theories}},
  \href{http://dx.doi.org/10.1016/S0370-2693(97)00530-3}{\emph{Phys. Lett. B}
  {\bfseries 404} (1997) 71--76},
  [\href{https://arxiv.org/abs/hep-th/9703215}{{\ttfamily hep-th/9703215}}].

\bibitem{Kapustin:2011gh}
A.~Kapustin, \emph{{Seiberg-like duality in three dimensions for orthogonal
  gauge groups}},  \href{https://arxiv.org/abs/1104.0466}{{\ttfamily
  1104.0466}}.

\bibitem{dancer2013implosion}
A.~Dancer, F.~Kirwan and A.~Swann, \emph{Implosion for hyperk{\"a}hler
  manifolds}, {\emph{Compositio Mathematica} {\bfseries 149} (2013)
  1592--1630}.

\bibitem{Kirwan2011}
F.~Kirwan, \emph{{Symplectic implosion and nonreductive quotients}},  in
  \emph{{Geometric aspects of analysis and mechanics}}, pp.~{213--256}, {
  Progr.Math. 292, Birkhauser/Springer. New York}, 2011.

\bibitem{CMcG}
D.~Collingwood and W.~McGovern, \emph{Nilpotent orbits in semisimple Lie
  algebras}.
\newblock van Nostrand Reinhold, 1993.

\bibitem{Nakajima:2018}
H.~Nakajima, \emph{{Introduction to a provisional mathematical definition of
  Coulomb branches of 3-dimensional N=4 gauge theories}},  in \emph{{Modern
  geometry: a celebration of the work of Simon Donaldson}}, pp.~193--211,
  {Proc. Sympos. Pure Math., 99, Amer. Math. Soc., Providence, RI}, 2018.

\bibitem{Boalch2008}
P.~Boalch, \emph{{Irregular connections and Kac-Moody root systems}},
  \href{https://arxiv.org/abs/0806.1050}{{\ttfamily 0806.1050}}.

\bibitem{Hanany:2016gbz}
A.~Hanany and R.~Kalveks, \emph{{Quiver Theories for Moduli Spaces of Classical
  Group Nilpotent Orbits}},
  \href{http://dx.doi.org/10.1007/JHEP06(2016)130}{\emph{JHEP} {\bfseries 06}
  (2016) 130}, [\href{https://arxiv.org/abs/1601.04020}{{\ttfamily
  1601.04020}}].

\bibitem{Bourget:2020xdz}
A.~Bourget, J.~F. Grimminger, A.~Hanany, R.~Kalveks, M.~Sperling and Z.~Zhong,
  \emph{{Magnetic Lattices for Orthosymplectic Quivers}},
  \href{http://dx.doi.org/10.1007/JHEP12(2020)092}{\emph{JHEP} {\bfseries 12}
  (2020) 092}, [\href{https://arxiv.org/abs/2007.04667}{{\ttfamily
  2007.04667}}].

\bibitem{Bourget:2020gzi}
A.~Bourget, J.~F. Grimminger, A.~Hanany, M.~Sperling and Z.~Zhong,
  \emph{{Magnetic Quivers from Brane Webs with O5 Planes}},
  \href{http://dx.doi.org/10.1007/JHEP07(2020)204}{\emph{JHEP} {\bfseries 07}
  (2020) 204}, [\href{https://arxiv.org/abs/2004.04082}{{\ttfamily
  2004.04082}}].

\end{thebibliography}\endgroup

\end{document}